\documentclass[prb,amsfonts,floatfix,superscriptaddress]{revtex4}

\ifx\pdfoutput\undefined
\usepackage[dvips]{graphicx}
\else
\usepackage[pdftex]{graphicx}

\usepackage[pdftex]{hyperref}
\fi
\usepackage{amsmath,amsfonts,amssymb,amscd,amsthm,epsf}
\usepackage{subfigure}
\usepackage{color}

\graphicspath{{figs/}}

\theoremstyle{remark}

\newcommand{\Rset}{{\mathbb R}}
\newcommand{\Tset}{{\mathbb T}}
\newcommand{\nc}{\newcommand}
\nc{\figref}[1]{Fig.~\ref{fig:#1}}
\nc{\figsref}[2]{Figs.~\ref{fig:#1}-\ref{fig:#2}}
\nc{\tabref}[1]{Table~\ref{tab:#1}}
\nc{\tabsref}[2]{Tables~\ref{tab:#1}-\ref{tab:#2}}
\nc{\secref}[1]{Sec.~\ref{sec:#1}}
\nc{\secsref}[2]{Sec.~\ref{sec:#1}-Sec.~\ref{sec:#2}}
\nc{\ssecref}[1]{Sec.~\ref{ssec:#1}}
\nc{\ssecsref}[2]{Sec.~\ref{ssec:#1}-Sec.~\ref{ssec:#2}}
\nc{\eqeqref}[1]{Eq.~\eqref{eq:#1}}
\nc{\eqseqref}[2]{Eqs.~\eqref{eq:#1}-\eqref{eq:#2}}
\nc{\thmref}[1]{Theorem~\ref{theo:#1}}
\nc{\thmsref}[2]{Theorem~\ref{theo:#1}-\ref{theo:#2}}
\nc{\rcite}[1]{Ref.~\onlinecite{#1}}
\nc{\rcites}[1]{Refs.~\onlinecite{#1}}
\nc{\qtq}[1]{{\qquad\text{#1}\qquad}}
\nc{\vect}[1]{\boldsymbol{#1}}
\definecolor{changed}{rgb}{0.3,0.3,0.3}
\nc{\changed}[1]{{\bf previous version: }{\color {changed} #1}}

\begin{document}

\title{Breakup of Shearless Meanders and ``Outer'' Tori in the Standard
 Nontwist Map}

\author{K.~Fuchss} \affiliation{Department of Physics and Institute
for Fusion Studies, The University of Texas at Austin, Austin, TX
78712}

\author{A.~Wurm} \affiliation{Department of Physical \& Biological
Sciences, Western New England College, Springfield, MA 01119}

\author{A.~Apte} \affiliation{Department of Mathematics, University of
North Carolina at Chapel Hill, Chapel Hill, NC 27599}

\author{P.J.~Morrison} \affiliation{Department of Physics and
Institute for Fusion Studies, The University of Texas at Austin,
Austin, TX 78712}

\date{\today}

\begin{abstract}
The breakup of shearless invariant tori with winding number
$\omega=[0,1,11,1,1,\ldots]$ (in continued fraction representation) of
the standard nontwist map is studied numerically using Greene's
residue criterion. Tori of this winding number can assume the shape of
meanders (folded-over invariant tori which are not graphs over the
$x$-axis in $(x,y)$ phase space), whose breakup is the first point of
focus here. Secondly, multiple shearless orbits of this winding number
can exist, leading to a new type of breakup scenario.  Results are
discussed within the framework of the renormalization group for
area-preserving maps. Regularity of the critical tori is also
investigated.
\end{abstract}

\maketitle

{\bf In recent years {\it nontwist maps}, area-preserving maps that
violate the twist condition locally in phase space, have been the
subject of several studies in physics and mathematics. These maps
appear naturally in a variety of physical models.  An important
problem is the understanding of the breakup of invariant tori, which
correspond to transport barriers in the physical model.  We conduct a
detailed study of the breakup of two types of invariant tori that have
not been analyzed before.}

\section{Introduction}
\label{sec:intro}

We consider the {\it standard nontwist map} (SNM) $M$ as introduced in
\rcite{del_castillo93},
\begin{eqnarray}
x_{n+1} & = & x_n + a \left( 1-y^2_{n+1}\right)\nonumber\\
y_{n+1} & = & y_n -b \sin\left(2\pi x_n\right)\,,
\end{eqnarray}
where $(x,y)\in \Tset\times \Rset$ are phase space coordinates and
$a,b\in\Rset$ are parameters. This map is area-preserving and violates
the {\it twist condition}, $\partial x_{n+1}(x_n,y_n)/\partial y_n
\neq 0$, along a curve in phase space. Although the SNM is not generic
due to its symmetries, it captures the essential features of nontwist
systems with a local, approximately quadratic extremum of the winding
number profile.

Nontwist maps are low-dimensional models of many physical systems, as
described in \rcites{wurm05,apte03,del_castillo96}.  Of particular
interest from a physics perspective is the breakup of invariant tori
(which we alternatively call invariant curves), consisting of
quasiperiodic orbits with irrational winding number (see Appendix),
that often correspond to transport barriers in the physical system.

One important characteristic of nontwist maps is the existence of
multiple orbit chains of the same winding number. For the SNM, in
particular, the symmetry $S(x,y)=(x+1/2,-y)$ guarantees that whenever
an orbit chain of a certain winding number exists, a second chain with
the same winding number can be found.

Changing the map parameters $a$ and $b$ causes bifurcations of
periodic orbit chains with the same winding number.  Orbits can
undergo stochastic layer reconnection (``separatrix'' reconnection),
or they can collide and annihilate. The simplest
reconnection-collision scenarios, which involve sequences of
collisions between elliptic and hyperbolic orbits of a single pair of
periodic orbit chains with the same winding number, have been known
since early study of nontwist systems.\cite{howard84} In the SNM, two
{\it standard} scenarios can be distinguished that describe,
respectively, the reconnection-collision sequence for a pair of either
even or odd-period orbit chains. A detailed review of these scenarios,
as well as a discussion of earlier studies of reconnection-collision
phenomena in theory and experiments, can be found in \rcite{wurm05}.

\begin{figure}[t!] \begin{center}
\includegraphics[width=3in]{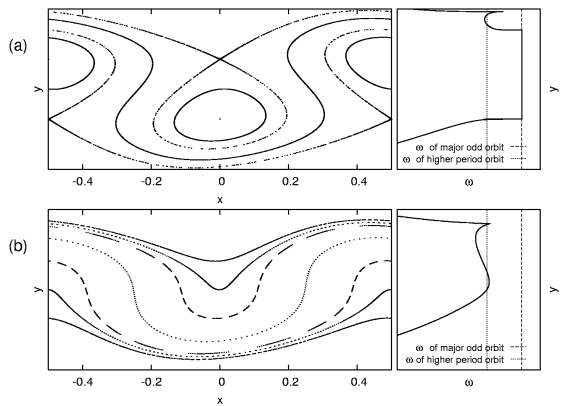}
\caption{Phase space (left) and winding number profile along $x=0.5$
  (right) of two stages of odd-period standard scenario. The
  $y$-ranges of all plots are identical. Parameters were chosen (a)
  slightly before collision and (b) slightly after collision of two
  major odd-period orbit chains (winding number marked by dashed
  lines).}
\label{fig:odd} \end{center} \end{figure}

As recently reported,\cite{wurm05} even the simple, non-generic SNM
can have more than two orbit chains of the same winding number, and
thus reconnection-collision scenarios are more intricate than
previously expected.  To illustrate this, two stages of the odd-period
standard scenario are shown in~\figref{odd}.  Each of the winding
number profiles exhibits two peaks at or somewhat below the winding
number of the colliding odd-period orbit, which is marked by the
dashed vertical line. Between the peaks lies a recess, i.e., for any
winding number between the lowest value of the recess and the peak
values, e.g., the value marked by dotted verticals, four (or more)
orbits of that winding number traverse the $x=0.5$ symmetry
line.\footnote{See Appendix for definition. For the usefulness of
symmetry lines in locating periodic orbits refer, e.g., to the
Appendix B of \rcite{wurm05} and references in the same paper.}
Similar observations can be made for other symmetry lines.  Therefore,
whenever a periodic orbit is studied with winding number inside the
recess associated with a major odd-period orbit collision, complicated
reconnection-collision scenarios are possible, which were called {\em
non-standard scenarios} in \rcite{wurm05}.

When two quasiperiodic orbits collide, the winding number profile
shows a local extremum and the orbit at collision is referred to as
the {\em shearless curve}. In previous studies of the SNM only the
shearless curves invariant under the full symmetry group $\cal G$ of
the SNM (composed of the symmetry $S$ as well as the involutions $I_1$
and $I_2$)\footnote{See Appendix B of \rcite{wurm05}.} have been
considered. However, as seen in \figref{odd}(b), in addition to the
central extremum (here a minimum) in the winding number profile, other
shearless curves (here marked by the two outer peaks) may
exist. Figure~\ref{fig:odd}(a) also exhibits peaks, but since the
plateau and spike are associated with an elliptic point and the
invariant manifolds of hyperbolic points, respectively, rather than
quasiperiodic orbits, they do not qualify as shearless {\em curves}.
From now on, we will refer to the $\cal G$-invariant curve as the
``central shearless curve'' and to others as ``outer shearless
curves.'' Since the breakup of outer shearless curves has not been
studied so far, this will be the focus of our investigation
in~\secref{outer}.

Another consequence of the violation of the twist condition is the
occurrence of {\em meanders}, quasiperiodic orbits that are ``folded
over'', i.e., not graphs $y(x)$. Whereas Birkhoff's theorem states
that such curves cannot exist in twist maps, they can occur in
nontwist maps. In the SNM, meanders appear between (in parameter
space) the reconnection and collision of odd-period
orbits.\cite{petrisor01} Figure~\ref{fig:odd}(a) shows an example,
whereas in Fig.~\ref{fig:odd}(b), the meander has changed to a graph
again. As seen in Fig.~\ref{fig:odd}(a), the region in which meanders
are found corresponds to a recess in the winding number profile.
However, the converse is not true: Fig.~\ref{fig:odd}(b) shows an
example where meanders are absent, but still a recess in the winding
number profile is observed. To our knowledge, the breakup of meanders
has not been studied previously.

The paper is organized as follows. In \secref{nontwist} we review
previous results about the SNM relevant to this investigation.
Section~\ref{ssec:nt-ovw} contains a brief discussion of the parameter
space of the SNM, in particular the details of where the scenarios
studied in this paper occur.  Section~\ref{ssec:breakup} contains an
account of how Greene's residue criterion is used for detecting the
breakup of critical invariant tori. In \secref{central} we discuss the
results for the breakup of the central shearless meander of winding
number $\omega=[0,1,11,1,1,\ldots]$ (in continued fraction
representation), while in \secref{outer} we consider the breakup of
the outer shearless tori of the same winding number. Questions of
regularity of these critical tori and a comparison with previous
results are addressed in \secref{regularity}. Finally,
\secref{conclusion} contains our conclusion and a discussion of open
questions. Basic definitions are given in the Appendix.

\section{Background}
\label{sec:nontwist}

\subsection{Parameter space overview}
\label{ssec:nt-ovw}

In order to identify the parameter regions where the bifurcations
described in \secref{intro} occur, various parameter space curves can
be computed (usually numerically). Collisions of periodic orbits are
described by the {\em bifurcation curves} \footnote{See the Appendix
for definitions.} introduced in \rcite{del_castillo96} and generalized
in \rcite{wurm05}.  Reconnections do not occur precisely on parameter
space curves (see, e.g., \rcite{apte05c}), but within a finite range
of parameters; however, the range is usually small enough that the
method of \rcite{petrisor02} (implemented in \rcite{wurm04}) yields
curves that represent a good approximation of the {\em reconnection
thresholds} for odd-period orbits. For even-period orbits,
reconnection coincides with the collision of hyperbolic orbits.

By numerically computing the {\em branching points} at which
bifurcation curves for various higher periodic orbits split up into
several branches below a major odd-period orbit collision, one obtains
a good estimate of the parameter region for which multiple orbit
chains exist. For an extensive discussion of these curves and their
computation see \rcite{wurm05}.

\begin{figure*}[t!] \begin{center}
\includegraphics[width=0.75\textwidth]{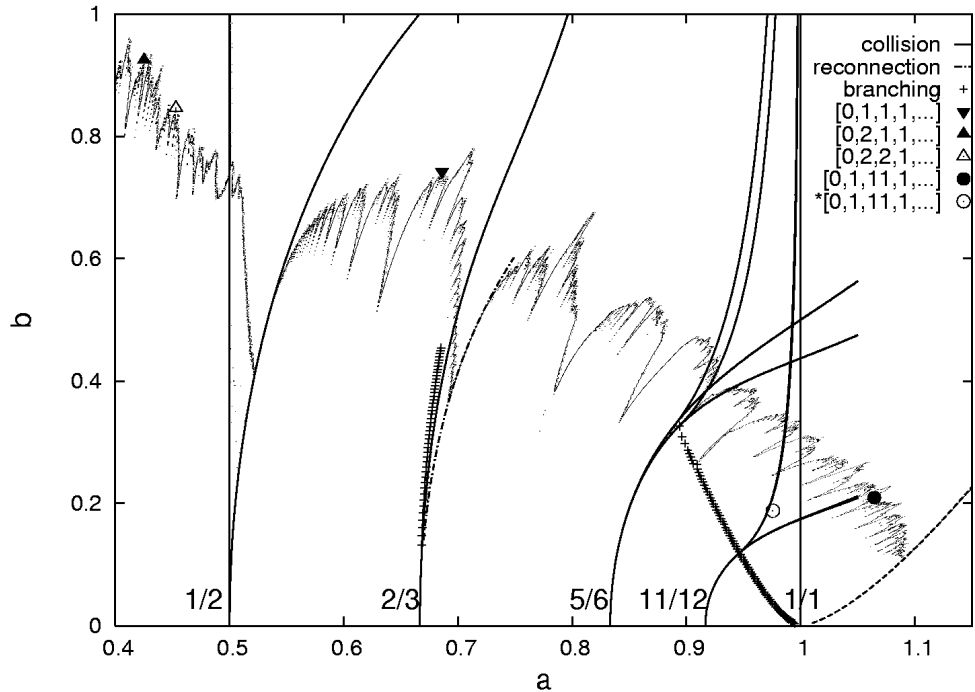}
\caption{Parameter space overview of SNM, with higher orbit branching
  points, bifurcation (collision) curves, and approximate reconnection
  thresholds for odd-period orbits 2/3 and 1/1.  Bifurcation curves
  for even-period orbits 1/2, 5/6, and 11/12 and the breakup boundary
  for the central shearless tori are indicated.  New points are marked
  by two circles. ($\odot$ denotes outer shearless curve, all others
  are central ones.)}
\label{fig:shinoplot} \end{center} \end{figure*}

An overview of parameter space with thresholds for several examples of
low-period orbits is shown in \figref{shinoplot}.  Higher orbit
branching points, bifurcation curves, and approximate reconnection
thresholds are shown for the odd-period orbits 2/3 and 1/1. Meanders
occur between reconnection and collision; a recess in winding number
profile is encountered in the region limited by branching points and
bifurcation curve. For even-period orbits with winding number 1/2,
5/6, and 11/12, only bifurcation curves are shown, since even-period
orbits do not induce branching and reconnections coincide with
collisions. Of the corresponding winding numbers, no orbits exist
above the highest (in $b$) bifurcation curve and two orbit chains are
found below the lowest one; in between various numbers of orbits exist
on each symmetry line.  In addition, the figure also contains the
ragged breakup boundary, introduced in \rcite{shinohara97}, above
which the central shearless orbits have become chaotic.\footnote{We
assume that an orbit is chaotic if its winding number is ``numerically
undefined'', i.e., within \mbox{3 000 000} iterations of an indicator point
on the shearless orbit, we could not find a sequence of \mbox{10 000}
consecutive approximations $\omega_i=x_i/i$ that ``converged'' without
assuming a global maximum or minimum within the sequence
($|\omega_i-\omega_{i-1}|<10^{-7}$, and $\omega_i<\max_{n<i}
\omega_n$, and $\omega_i>\min_{n<i} \omega_n$).}  We further indicate
points (by triangles) for which the breakup of the central shearless
curve has been studied in detail in the past (see
\rcites{del_castillo96,wurmdiss,apte03,apte05a,aptediss}) as well as
the two points investigated in this paper: The central meander from
\secref{central} is shown as a solid circle ($\bullet$), the outer
shearless curves from \secref{outer} as an empty circle ($\odot$).
Note that all breakup points for central shearless tori are located on
the breakup boundary, whereas the outer shearless curves break up at
smaller parameter values.

The winding number investigated here was chosen such that the central
shearless curve is a meander, i.e., the point $\bullet$ is located
between the 1/1 reconnection and collision thresholds, and that
multiple orbit chains and hence multiple shearless curves can occur,
i.e., the point $\odot$ is located to the right of the 1/1 branching
threshold. This can be ensured by picking a winding number close to a
periodic orbit, here $11/12$, whose bifurcation curve both branches
due to a nearby major odd-period orbit, and has one branch crossing a
collision threshold of this major odd-period orbit before crossing the
breakup boundary.

\subsection{Greene's residue criterion} \label{ssec:breakup}

Whereas the breakup boundary in \figref{shinoplot} provides a rough
estimate of the parameter values at which central shearless tori
break, a significantly more precise tool for studying the breakup of a
particular torus with given winding number is provided by Greene's
residue criterion, originally introduced in the context of twist
maps.\cite{greene79} This method relies on the numerical observation
that the breakup of an invariant torus with irrational winding number
$\omega$ is determined by the stability of nearby periodic
orbits. Some aspects of the validity of this criterion have been
proved for nontwist maps.\cite{delshams00}

To study the breakup, one considers a sequence of periodic orbits with
winding numbers $m_i/n_i$ converging to $\omega$,
$\lim_{i\rightarrow\infty} m_i/n_i=\omega$. The sequence converging
the fastest, and hence the most commonly used one, consists of the
convergents of the continued fraction expansion of $\omega$, i.e.,
$m_i/n_i=[a_0, a_1, \ldots, a_i]$, where
\begin{equation}
\label{eq:cf}
\omega=[a_0, a_1, a_2, \ldots] =
  a_0 + \cfrac{1}{a_1 + \cfrac{1}{a_2 + \ldots}}.
\end{equation}
The stability of the corresponding orbits can be expressed through
their residues, $R_i= [2-{\rm Tr}(DM^{n_i})]/4$, where ${\rm Tr}$ is
the trace and $DM^{n_i}$ is the linearization of the $n_i$ times
iterated map about the periodic orbit: An orbit is elliptic for
$0<R_i<1$, parabolic for $R_i=0$ and $R_i=1$, and hyperbolic
otherwise. The convergence or divergence of the residue sequence
associated with the chosen periodic orbit sequence then determines
whether the torus exists or not, respectively:
\begin{itemize}
\item $\lim_{i\rightarrow\infty}|R_i|=0$ if the torus of winding
  number $\omega$ exists.
\item $\lim_{i\rightarrow\infty}|R_i|=\infty$ if the torus of winding
  number $\omega$ is destroyed.
\end{itemize}
At the breakup itself, various scenarios can be encountered, depending
on the class of maps and invariant torus under consideration.

For twist systems, this criterion has been used to study the breakup
of noble invariant tori in the standard (twist) map, i.e., orbits with
winding numbers that have a continued fraction expansion tail of 1's
(see, e.g., \rcites{greene79,mackay82,mackay83}). It was found that at
the point of breakup the sequence of residues converges to either
$R_\infty\approx 0.25$ or a three-cycle containing $0.25\ldots$ as one
of its elements.

In the standard nontwist map, the residue criterion was first used in
\rcite{del_castillo96} to study the breakup of the central shearless
torus of inverse golden mean $1/\gamma = (\sqrt 5 -1)/2 =
[0,1,1,1,\ldots]$ winding number.  There it was discovered that the
residue sequence converges to a six-cycle.\footnote{It should be
noted, however, that in the nontwist case only orbits corresponding to
half of the continued fraction convergents exist in the parameter
space region of interest. Therefore the six-cycle corresponds to a
twelve-cycle when correctly compared with the twist case.}  Similar
studies were conducted for noble central shearless tori of winding
numbers $\omega=1/\gamma^2$ (\rcites{wurmdiss,apte03}) and
$\omega=[0,2,2,1,1,1,\ldots]$ (\rcite{aptediss}), and the same
six-cycle was found. The parameter values at which these shearless
tori break, i.e., at which six-cycles of residues are encountered, are
marked by triangles in \figref{shinoplot}. In this paper, we study the
noble winding number $\omega=[0,1,11,1,1,1,\ldots]$, where the large
number 11 in the second convergent had to be chosen to ensure that the
breakup occurs in a region in parameter space where both meanders and
multiple shearless tori are possible, as described in
\ssecref{nt-ovw}.

In addition to nontrivial residue convergence behavior, invariant tori
at breakup exhibit scale invariance under specific phase space
re-scalings. All these results suggest that certain characteristics of
the breakup of noble invariant tori are universal, i.e., the same
within a large class of area-preserving maps. To interpret the
results, a renormalization group framework based on the residue
criterion has been developed (see, e.g., \rcites{mackay83,
del_castillo97,apte03,apte05a}).

\section{Breakup of the $\omega=[0,1,11,1,1,\ldots]$ central shearless
  meander}
\label{sec:central}

\subsection{Search for critical parameter values}
\label{ssec:cf}

In order to study a shearless irrational orbit, one needs to locate
parameter values on its bifurcation curve. This can be achieved
numerically by approximating them by parameter values on the
bifurcation curves of nearby periodic orbits, usually of orbits with
winding numbers that are the continued fraction convergents of
$\omega$. For $\omega=[0,1,11,1,1,\ldots]\approx
0.920748351059159504$, the convergents up to the highest numerically
accessible one in our studies are shown in~\tabref{cfs}.

\begin{table}[t!]
\begin{center}
\begin{tabular}{|r|l||r|l||r|l|} \hline
$n$ & $[n]$ & $n$ & $[n]$ & $n$ & $[n]$ \\
\hline \hline
0 & 1/1 & 13 & 2707/2940 & 26 & 1410348/1531741 \\
1 & 11/12 & 14 & 4380/4757 & 27 & 2281991/2478409 \\
2 & 12/13 & 15 & 7087/7697 & 28 & 3692339/4010150 \\
3 & 23/25 & 16 & 11467/12454 & 29 & 5974330/6488559 \\
4 & 35/38 & 17 & 18554/20151 & 30 & 9666669/10498709 \\
5 & 58/63 & 18 & 30021/32 605 & 31 & 15640999/16987268 \\
6 & 93/101 & 19 & 48575/52756 & 32 & 25307668/27485977 \\
7 & 151/164 & 20 & 78596/85361 & 33 & 40948667/44473245 \\
8 & 244/2 65 & 21 & 127171/138117 & 34 & 66256335/71959222 \\
9 & 395/429 & 22 & 205767/223478 & 35 & 107205002/116432467 \\
10 & 639/694 & 23 & 332938/361595 & 36 & 173461337/188391689 \\
11 & 1034/1123 & 24 & 538705/585073 & 37 & 280666339/304824156 \\
12 & 1673/1817 & 25 & 871643/946668 & 38 & 454127676/493215845 \\
\hline
\end{tabular}
\end{center}
\caption{Continued fraction convergents for $[0,1,11,1,1,\ldots]$,
  where $[n]=[0,a_0,\ldots,a_{n+2}]$ (following the notation
  of~\rcite{apte03}).}
\label{tab:cfs}
\end{table}

\begin{figure}[t!] \begin{center}
\includegraphics[width=3in]{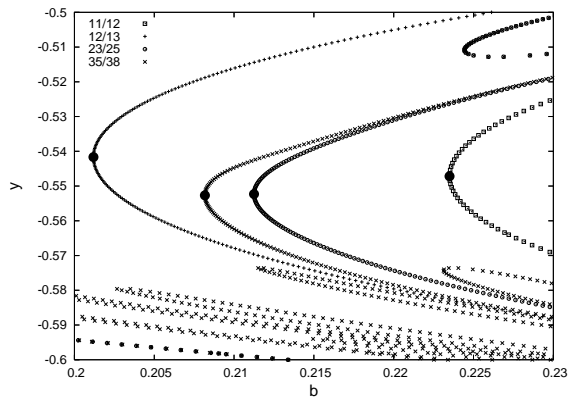}
\caption{ $y$-values in phase space of the 11/12, 12/13, 23/25, and
  35/38 central orbits on the $s_1$ symmetry line as function of $b$,
  at $a=1.0645342893$.}
\label{fig:rootby} \end{center} \end{figure}

For given parameters $(a,b)$, any of these periodic orbits (if they
exist) can be found along symmetry lines via a one-dimensional root
search, as explained, e.g., in \rcite{del_castillo96}. Performing this
search for a range of parameters, usually varying $b$ while keeping
$a$ constant, results in the relation $y(b)$, i.e., the location(s) of
the periodic orbit along a given symmetry line, as shown in
Fig.~\ref{fig:rootby} for the orbits 11/12, 12/13, 23/25, and 35/38
along the $s_1$ symmetry line.

In this plot, orbit collisions are found where two $y(b)$ branches
meet, i.e., at extrema of $b(y)$. Especially for higher period orbits,
multiple collisions can be observed, but in this section we focus only
on the ones approximating the central shearless curve, deferring the
primary outer ones (i.e., the only additional collisions found for the
lowest, 11/12, orbit) to~\secref{outer}. In contrast to previous
publications, in which central collisions appear as maxima in $b(y)$,
here, in the meander regime, they are associated with minima.

The parameter values $b_{[n]}(a)$ of these central collisions, found
by extremum searches and marked by solid circles ($\bullet$) in
Fig.~\ref{fig:rootby}, converge to $b_\infty(a)$ (located on the
bifurcation curve of $\omega$). Now Greene's residue criterion, as
described in \ssecref{breakup}, can be used to determine whether at
$\left(a,b_\infty(a)\right)$, the shearless curve still exists or not:
At parameter values $a$ and the best known approximation to
$b_\infty(a)$, the residues of all periodic orbits of convergents that
have not collided, here the orbits $[n]$ with even $n$, are
computed. Their limiting behavior for $n\rightarrow\infty$ reveals the
status of the torus. By repeating the procedure for various values of
$a$, with alternating residue convergence to 0 and $\infty$, the
parameter values of the shearless torus breakup, $(a_c, b_\infty(a_c)
)$, can be determined to high precision.

\begin{figure}[t!] \begin{center}
\includegraphics[width=3in]{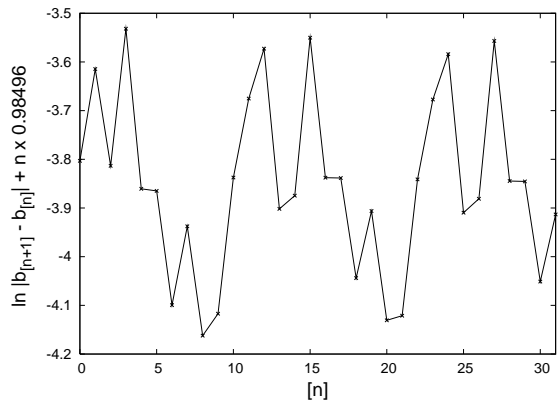}
\caption{12-cycle of $b_{[n]}$ differences in approximating the
  critical shearless $[0,1,11,1,1,\ldots]$ meander at
  $a_c=1.0645342893$ on $s_1$.
  \label{fig:bdiffmeander}}
\end{center} \end{figure}

Due to numerical limitations, the highest orbit collision used here to
approximate $b_\infty(a)$ is $b_{[33]}(a)$.  However, a better
approximation can be obtained by observing (in hindsight) that close
to the critical breakup value, the $b_{[n]}(a)$ obey a scaling law
\begin{equation}
\label{eq:bscalinglaw}
b_{[n]}=b_\infty + B(n) \delta_1 ^{-n} \;,
\end{equation}
where $B(n)$ is empirically found to be periodic in $n$ with period 12
as $n\rightarrow \infty$. As $b_\infty$ is unknown, the scaling is
most readily observed by plotting
\begin{equation}
\label{eq:bscalinglog} \ln \left( b_{[n+1]}-b_{[n]} \right) =
\tilde{B}(n) -n\ln \delta_1 \;,
\end{equation}
where $\tilde{B}(n)=\ln(B(n+1)/\delta_1 -B(n))$ is also periodic in
$n$ with period 12. This is shown in \figref{bdiffmeander}, where for
clarity only the offsets of $\ln \left( b_{[n+1]}-b_{[n]} \right)$
about the average slope are shown. The $b_{[n]}$ values used here were
obtained from orbits colliding on the $s_1$ symmetry line, although
the same behavior is observed on the $s_2$ symmetry line. For $s_3$
and $s_4$, a similar plot is found, however, the 12-cycles are shifted
by $n\pm 6$.  The slope was calculated from the last 24 difference
values by averaging the last 12 slopes $\left(
\ln(b_{[n+13]}-b_{[n+12]})- \ln(b_{[n+1]}-b_{[n]}) \right) /12$, with
$n=8,\ldots,20$. The result is $\log\delta_1 \approx 0.98496\pm
0.00036$, or
\begin{equation}
\label{eq:delta1a} \delta_1=2.678 \pm 0.001 \:.
\end{equation}
The periodicity of $\tilde{B}(n)$ enables us to obtain a better
approximation, $b^*$ (i.e., closer to $b_\infty$ than $b_{[33]}$) from
lower $b_{[n]}$-values, using the extrapolation
\begin{equation}
\label{eq:extrapolation}
b^* = b_{[32]}+
 \frac{b_{[32]}-b_{[20]}}{(b_{[21]}-b_{[20]})-(b_{[33]}-b_{[32]})}
 \,\times\,(b_{[33]}-b_{[32]})
\end{equation}
for the best shearless $b_\infty(a)$-approximation at which to apply
Greene's residue criterion.

%

\subsection{Residue six-cycle at breakup} \label{ssec:sixc}

\begin{figure}[t!] \begin{center}
\includegraphics[width=0.45\textwidth]{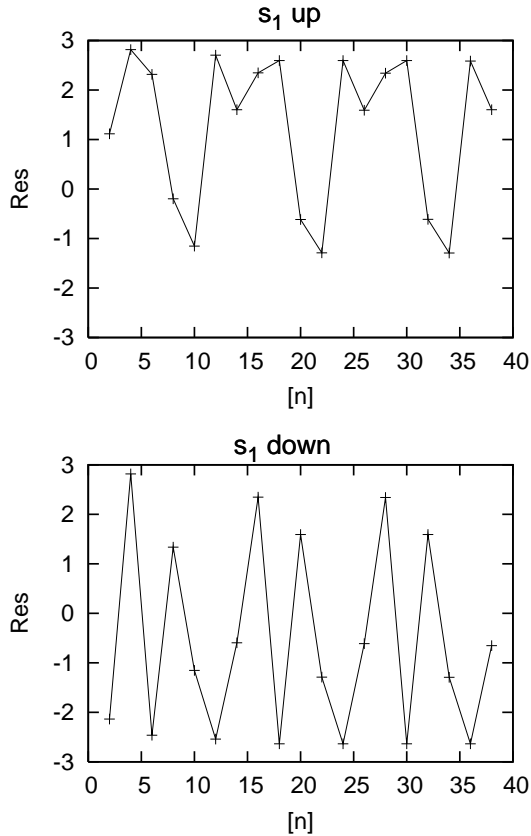}
\caption{Residue six-cycles for orbits $s_1$ at meander
  breakup. Cycles for orbits on $s_2$ are the same, with up and down
  interchanged. Cycles on $s_3$ are shifted by three, with up and down
  interchanged. Cycles on $s_4$ are shifted by three. }
\label{fig:resm6c} \end{center} \end{figure}

Searching along $\left( a,b^*(a)\right)$ for the transition between
residue convergence to 0 and $\infty$, we obtain as the critical
parameters for the shearless meander breakup:
\begin{equation}
\label{eq:acbc}
(a_c, b_c)=(1.0645342893, 0.209408148327230359)\;.
\end{equation}
At these parameters, only orbits with $n$ even in \tabref{cfs} exist,
two for each $n$, denoted as ``up'' and ``down'' for larger and
smaller $y$-values on a symmetry line, respectively.  Plotting the
residues of these orbits, one observes the 6-cycles in
\figref{resm6c}, here shown for the up and down orbits on $s_1$.

The same cycles are found for the other symmetry lines, with up and
down orbits interchanged and shifted by $[n] \rightarrow [n+6]$, as
summarized in \tabref{6cscheme}. Since these cycles are the same as
the ones observed in~\rcite{apte03} (up to an interchange of up and
down orbits and a shift of $[n] \rightarrow [n+7]$), the labels $C_i$
and $D_i$ here were assigned to reflect this correspondence.

\begin{table}[t!]
\begin{center}
\begin{tabular}{|r||c|c|c|c|} \hline
\multicolumn{1}{|c||}{$[n]$} & $R_{u1}=R_{d2}$ & $R_{u2}=R_{d1}$ &
$R_{u3}=R_{d4}$ & $R_{u4}=R_{d3}$ \\
\hline \hline
$[8], [20], [32]$ & $C_1$ & $D_1$ & $C_4$ & $D_4$\\
$[10], [22], [34]$ & $C_2$ & $D_2$ & $C_5$ & $D_5$\\
$[12], [24], [36]$ & $C_3$ & $D_3$ & $C_6$ & $D_6$\\
$[2], [14], [26]$ & $C_4$ & $D_4$ & $C_1$ & $D_1$\\
$[4], [16], [28]$ & $C_5$ & $D_5$ & $C_2$ & $D_2$\\
$[6], [18], [30]$ & $C_6$ & $D_6$ & $C_3$ & $D_3$\\
\hline
\end{tabular}
\end{center}
\caption{Period-six convergence pattern of residues near criticality
  on different symmetry lines (following the notation
  of~\rcite{apte03}). Symmetry properties of the SNM further imply
  $C_6=C_3$, $D_1=C_4$, $D_2=C_2$, $D_4=C_1$, $D_5=C_2$, $D_6=D_3$.}
\label{tab:6cscheme}
\end{table}

\begin{figure*}[t!] \begin{center}
\includegraphics[width=0.8\textwidth]{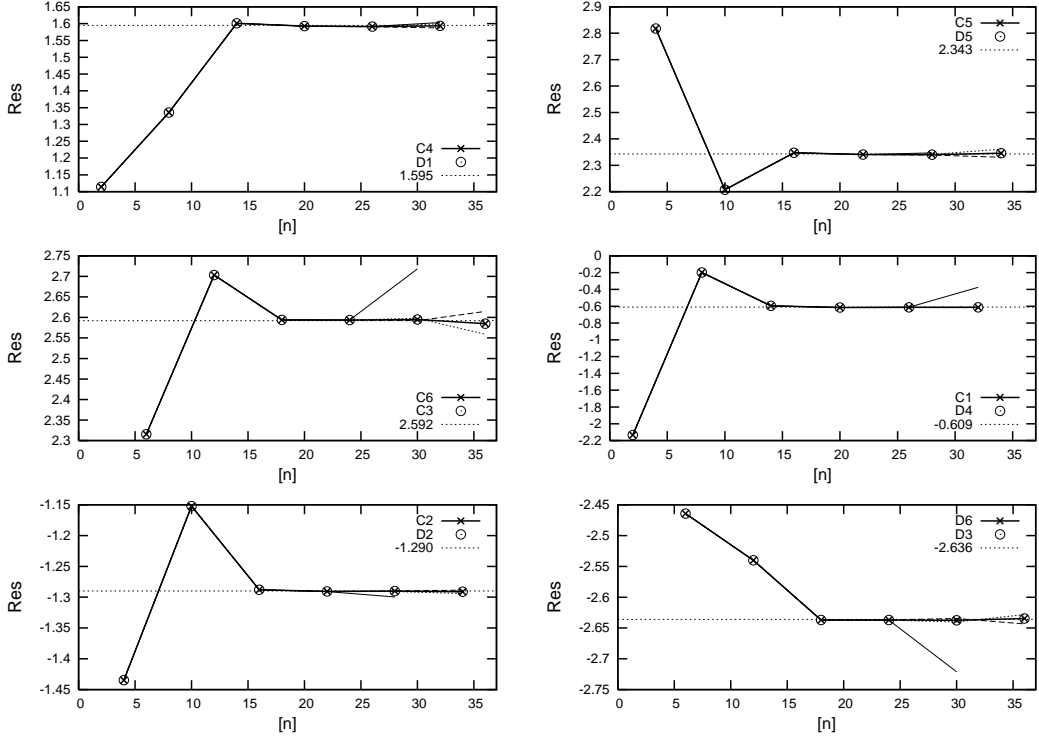}
\caption{Convergence of residue values at meander breakup. Shown are
  the two corresponding pairs of residues at the critical parameter
  values $(a_c, b_c)$ (bold line), at $(a_-, b_-)$ (dashed line), and
  $(a_+, b_+)$ (dotted line), all using extrapolated value
  $b^{\ast}(a)$ from the bifurcation curve $b_{[n]}$values of the
  [20], [21], [32], and [33] orbits. The thin line shows residue
  behavior without extrapolation for $\left( a_c, b_{33}(a_c)
  \right)$.}
\label{fig:resmc} \end{center} \end{figure*}

The convergence of the twelve individual values of the six-cycles, of
which due to symmetry properties of the map only six are truly
distinct as listed in \tabref{6cscheme}, are shown in \figref{resmc}.
\tabref{6cvalues} gives the corresponding numerical values. In
addition to the residues at the critical parameters for the shearless
meander breakup, at $a_c= 1.06453428930$ and
$b_c=0.209408148327230359$ (bold), residues slightly below, at
$a_-=1.06453428925$ and $b_-=0.209408148282494088$ (dashed), and
slightly above, at $a_+=1.06453428935$ and $b_+ =
0.209408148371966630$ (dotted) are displayed.  All these parameters
are extrapolated values from the bifurcation curves of the [20], [21],
[32], and [33] orbits. For comparison, the thin solid line shows
residue behavior without extrapolation, for $a_c=1.06453428930$ and
$b_{33}(a_c)=0.209408148327230605$ at the [33] orbit collision.

In summary, the six independent residues are
\begin{equation*}
\begin{array}{lrlr} C_1 =& -0.6090 \pm 0.0046, & C_2 =& -1.2901 \pm
0.0007,\\ C_4 =& 1.5945 \pm 0.0022, & C_5 =& 2.3434 \pm 0.0018,\\
C_6 =& 2.5919 \pm 0.0023, & D_6 =& -2.6365 \pm 0.0007,
\end{array}
\end{equation*}
where these numerical values were calculated each as the average of
the last four corresponding values in \tabref{6cvalues}, and the error
given here is the standard error (standard deviation of the mean).

\begin{table*}[t!]
\begin{center}
\begin{tabular}{|c||c|c|c|c||c|c|c|c|c|} \hline
$[n]$ & & $(a_-,b_-)$ & $(a_c,b_c)$ & $(a_+,b_+)$
& & $(a_-,b_-)$ & $(a_c,b_c)$ & $(a_+,b_+)$\\
\hline \hline
$[2]$ & $C_1$ & -2.1346 & -2.1346 & -2.1346 &
$C_4$ & 1.1143 & 1.1143 & 1.1143\\
$[8]$ & & -0.1976 & -0.1976 & -0.1976 &
& 1.3358 & 1.3358 & 1.3358\\
$[14]$ & & -0.5954 & -0.5954 & -0.5954 &
& 1.6009 & 1.6009 & 1.6009\\
$[20]$ & & -0.6151 & -0.6151 & -0.6151 &
& 1.5925 & 1.5925 & 1.5925\\
$[26]$ & & -0.6125 & -0.6126 & -0.6127 &
& 1.5909 & 1.5913 & 1.5917\\
$[32]$ & & -0.6114 & -0.6130 & -0.6145 &
& 1.5873 & 1.5934 & 1.5995\\
\hline
$[4]$ & $C_2$ & -1.4346 & -1.4346 & -1.4346 &
$C_5$ & 2.8180 & 2.8180 & 2.8180\\
$[10]$ & & -1.1523 & -1.1523 & -1.1523 &
& 2.2076 & 2.2076 & 2.2077\\
$[16]$ & & -1.2880 & -1.2880 & -1.2880 &
& 2.3475 & 2.3475 & 2.3475\\
$[22]$ & & -1.2907 & -1.2908 & -1.2908 &
& 2.3405 & 2.3406 & 2.3406\\
$[28]$ & & -1.2897 & -1.2903 & -1.2908 &
& 2.3389 & 2.3402 & 2.3414\\
$[34]$ & & -1.2886 & -1.2911 & -1.2940 &
& 2.3303 & 2.3455 & 2.3612\\
\hline
$[6]$ & $D_6$ & -2.4643 & -2.4643 & -2.4643 &
$C_6$ & 2.3155 & 2.3155 & 2.3155\\
$[12]$ & & -2.5402 & -2.5402 & -2.5402 &
& 2.7032 & 2.7032 & 2.7032\\
$[18]$ & & -2.6371 & -2.6371 & -2.6372 &
& 2.5938 & 2.5939 & 2.5939\\
$[24]$ & & -2.6368 & -2.6370 & -2.6372 &
& 2.5935 & 2.5937 & 2.5939\\
$[30]$ & & -2.6346 & -2.6374 & -2.6402 &
& 2.5921 & 2.5950 & 2.5980\\
$[36]$ & & -2.6434 & -2.6346 & -2.6284 &
& 2.6148 & 2.5850 & 2.5589\\
\hline
\end{tabular}
\end{center}
\caption{Numerical values of the six independent residues.}
\label{tab:6cvalues}
\end{table*}
%

\subsection{Phase space scaling invariance and renormalization
  results}
\label{ssec:scaling}

\begin{figure}[t!] \begin{center}
\includegraphics[width=3in]{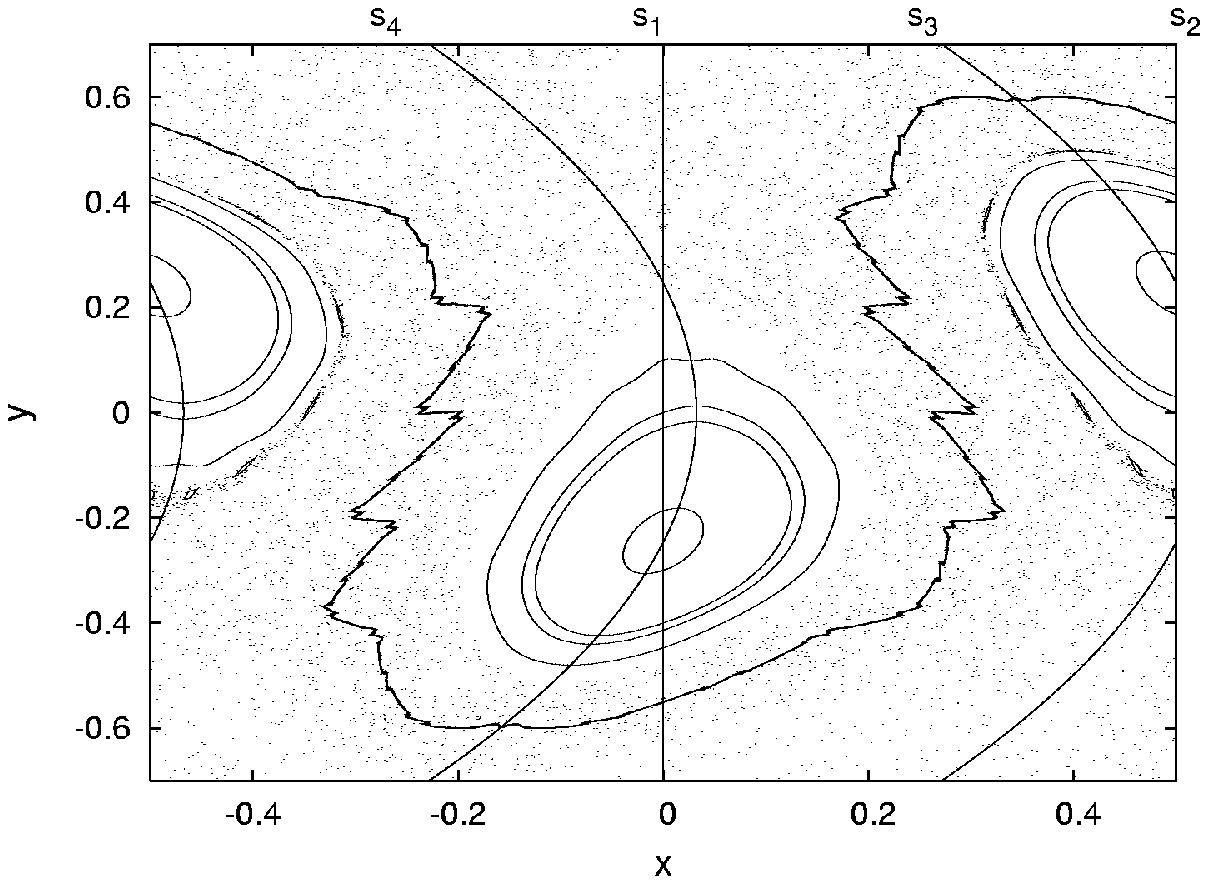}
\caption{ Phase space at the $[0,1,11,1,1,1\ldots]$ meander breakup,
  $a_c=1.0645342893$ and $b_c=0.209408148327230359$.  Also shown are
  the symmetry lines $s_1$, $s_2$, $s_3$, and $s_4$.}
\label{fig:meander-phsp} \end{center} \end{figure}

The phase space at the critical parameter values for the meander
breakup is shown in \figref{meander-phsp}. As in previous studies, the
shearless meander at breakup is scale invariant under specific
re-scalings of phase space

This is readily seen by zooming in at a certain point using different
levels of magnification: For example, we zoom in on the intersection
$(x_s,y_s)$ of the shearless meander with the $s_3$ symmetry line, and
transform to {\em symmetry line coordinates}, in which the $s_3$
symmetry line becomes a straight line,
\begin{eqnarray}
\label{eq:sln-coords}
x' &=& x-\frac{a}{2}\:(1-y^2) \:, \nonumber\\
y' &=& y-y_s \:.
\end{eqnarray}
Here, $y_s=0.59790858154$ was obtained by applying the same scaling as
in \eqeqref{extrapolation} for $b_{[n]}$ values to the $y_{[n]}$
locations of the $[22]$, $[24]$, $[34]$, and $[36]$ periodic orbits on
$s_3$, once for the respective up orbits and once for the down orbits,
and averaging over both results.

\begin{figure}[t!] \begin{center}
\includegraphics[width=3in]{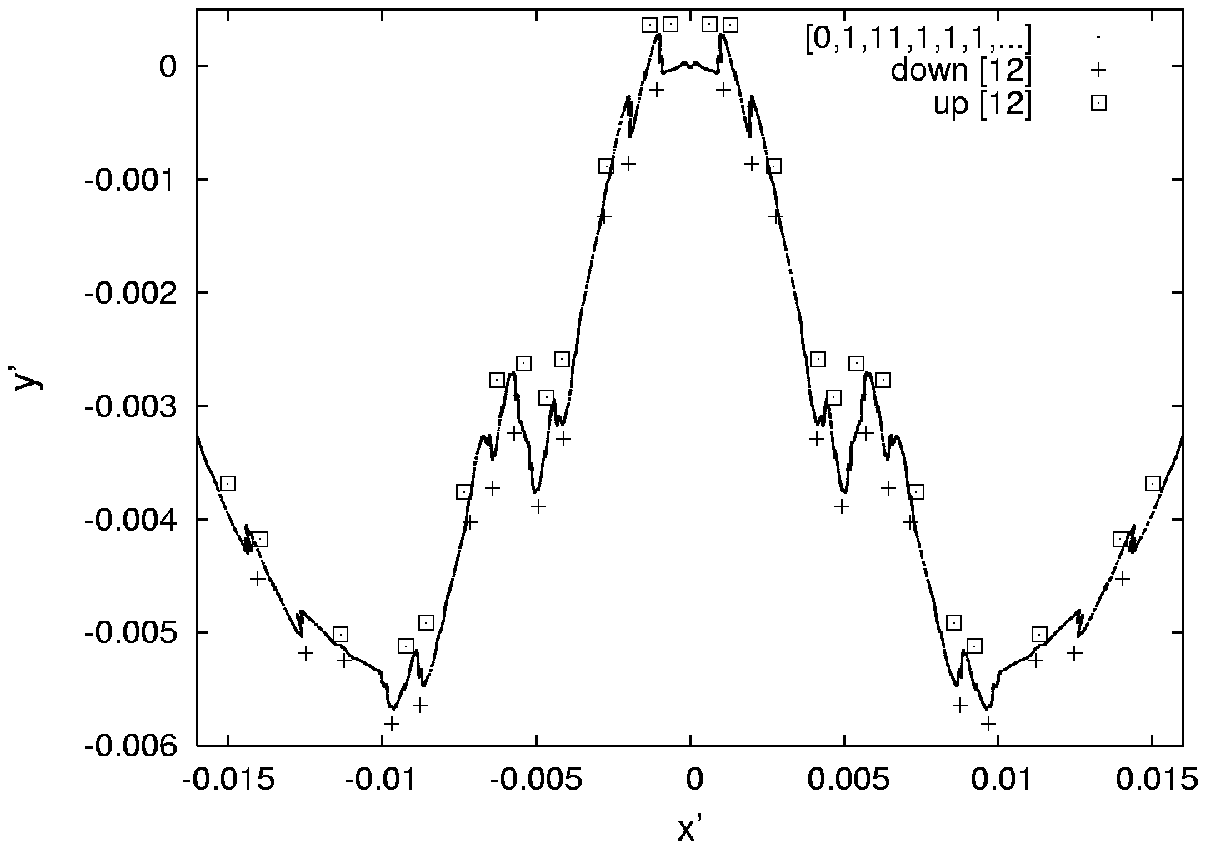}\\
\includegraphics[width=3in]{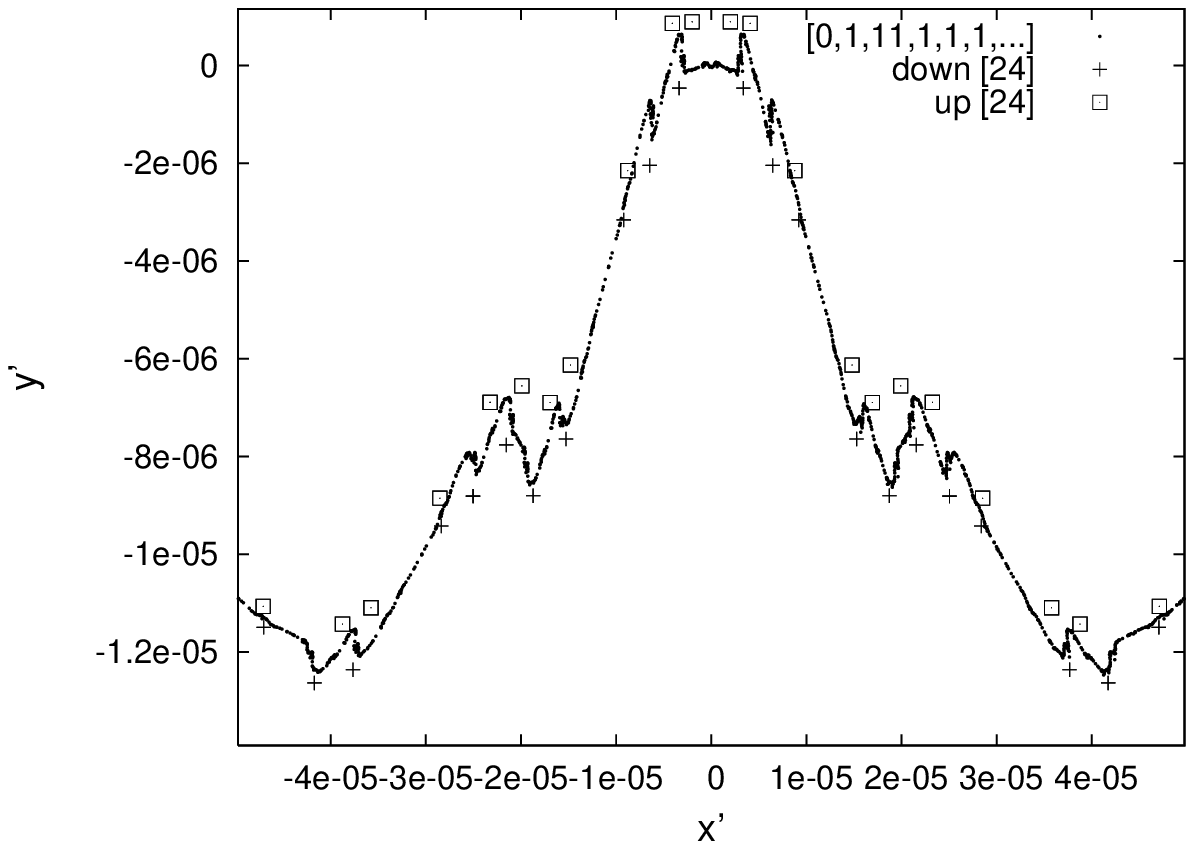}
\caption{Two levels of magnification, in symmetry line coordinates
  ($x',y'$), of the $[0,1,11,1,1,1\ldots]$ meander at breakup.  Also
  shown are the nearby up and down orbits of the $[12]$\/th (top) and
  $[24]$\/th (bottom) continued fraction convergents.}
\label{fig:meander-phsp-zoom} \end{center} \end{figure}

Figure~\ref{fig:meander-phsp-zoom} shows two levels of magnification
of the meander in these coordinates, each along with the up and down
periodic orbits of one of its convergents. The plotted region was
chosen to allow a direct comparison with Fig.~7 of \rcite{apte03}.
Although the two plots deviate slightly from each other towards the
edges of the plotted regions (because in contrast to \rcite{apte03}
the $x'$ and $y'$ ranges are larger here, i.e., scales at which the
meander is still influenced mostly by lower convergents), they
correspond exactly around the origin.

For a quantitative analysis, we compute the scaling factors $\alpha$
and $\beta$ such that the meander in the vicinity of its intersection
with $s_3$ is invariant under $(x',y') \rightarrow (\alpha^{12} x',
\beta^{12} y')$ (following the notation of \rcites{del_castillo97,
apte03}). Again, these are found from the limiting behavior of
convergent periodic orbits: Denoting by $(\hat{x}'_{n,\pm},
\hat{y}'_{n,\pm})$ the symmetry line coordinates of the point on the
up ($+$) or down ($-$) orbit of the $[n]$\/th convergent that is
located closest to $(0,0)$, we compute
\begin{equation}
\alpha^{12}_{n,\pm} =
   \left| \frac{\hat{x}'_{n,\pm}}{\hat{x}'_{n+12,\pm}} \right|
   \:, \:\:\:
\beta^{12}_{n,\pm} =
   \left| \frac{\hat{y}'_{n,\pm}}{\hat{y}'_{n+12,\pm}} \right|
   \:.
\label{eq:alphabeta}
\end{equation}
Averaging the six values $\alpha^{12}_{18\pm}$, $\alpha^{12}_{20\pm}$,
and $\alpha^{12}_{22\pm}$, we find $\alpha^{12}=321.65\pm 0.070$,
i.e., $\alpha=1.61789 \pm 0.00003$ (with the error being the standard
deviation of the mean).  Similarly, for $\beta$, we obtain
$\beta^{12}=431.29 \pm 0.19$, i.e., $\beta= 1.65792 \pm
0.00006$. These are the scaling factors used in
\figref{meander-phsp-zoom}. Within numerical accuracy, they coincide
with the values found in \rcites{del_castillo97, apte03}.

To interpret the scaling invariance of the shearless meander itself
and its convergents under $[n] \rightarrow [n+12]$ one can introduce a
renormalization picture, with a renormalization group operator ${\cal
R}_\omega$ acting on the space of maps with shearless curve at winding
number $\omega$ (see \rcites{mackay83, del_castillo97,
apte05a}). Operating with ${\cal R}_{[0,1,11,1,1,\ldots]}$ infinitely
many times on the standard nontwist map at criticality of the
shearless meander studied here limits to a map that is a period-12
fixed point, $\Lambda$, of the renormalization group operator.

In the vicinity of $\Lambda$, two unstable eigenvalues $\delta_1$ and
$\delta_2$ can be computed to characterize the fixed point. As shown
in \rcite{del_castillo97}, these are given by
$\delta_i=\sqrt[12]{1/\nu_i}$, where
\begin{eqnarray}
\nu_1&=& \lim_{n\rightarrow \infty} \left( \frac{b_{[n+12]}-b_c} {b_{[n]}-b_c}
\right) \:,\nonumber\\
\nu_2&=& \lim_{n\rightarrow \infty} \left( \frac{a_{c[n+12]}-a_c}
   {a_{c[n]}-a_c} \right) \:,
\label{eq:eigenvalues}
\end{eqnarray}
where $a_{c[n]}$ is the $a$-value along the bifurcation curve of the
$[n]$\/th convergent, rather than along the shearless curve, at which
the sequence of convergent residues exhibits nontrivial limiting
behavior (i.e., converges neither to 0 nor $\infty$). Using $n=21$, we
obtain the eigenvalues
\begin{equation}
\delta_1 \approx 2.680\:, \:\:\:
\delta_2 \approx 1.584,
\label{eq:eigenvalues-num}
\end{equation}
which are, within a small numerical error, the same values that were found for
the $1/\gamma^2$ shearless curve in \rcite{apte03}, and, with a slightly
larger error, for $1/\gamma$ in \rcite{del_castillo97}.\footnote{To get an
estimate of the accuracy of $\delta_1$, we computed it again for $n=20$: The
value was $2.689$, however, the true deviation before rounding was less that
$10^{-3}$. For $\delta_2$, the error is slightly harder to estimate, since the
$a$-values of nontrivial residue behavior are similarly hard to identify
numerically as for the outer shearless curves in \figref{reso1234}. Estimating
from similar figures that $a_{c[21]}$ is known with an accuracy of $10^{-10}$
and $a_{c[21]}$ with an accuracy of $2\times 10^{-9}$, and computing
$\delta_2$ again with minimal and maximal values results in
$\delta_{2-}=1.571$ and $\delta_{2+}=1.598$ as approximate outer bounds.}

\section{Breakup of the $\omega=[0,1,11,1,1,\ldots]$ outer shearless
  tori}
\label{sec:outer}

\subsection{Search for critical parameter values}
\label{ssec:cfouter}

\begin{figure}[t!] \begin{center}
\includegraphics[width=3in]{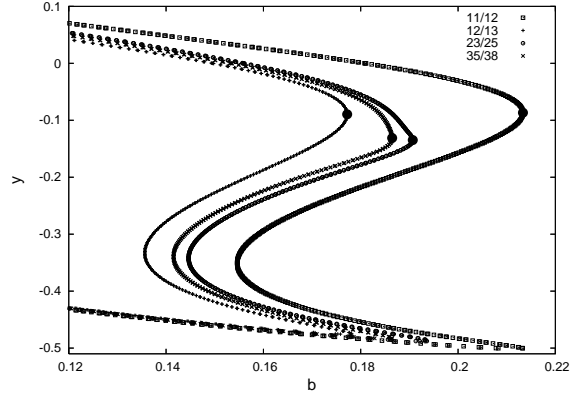}
\caption{ $y$-values in phase space of the 11/12, 12/13, 23/25, and
  35/38 outer orbits on the $s_1$ symmetry line as a function of $b$,
  at $a=0.9757564461$.}
\label{fig:rootbyouter} \end{center} \end{figure}

The procedure for finding the critical parameter values for the
breakup of outer tori is the same as that described in \ssecref{cf}.
The relations $y(b)$, at fixed $a$, for each of these orbits from
\tabref{cfs} are found along symmetry lines. The collision parameters
$b_{[n]}$ are again extrema of $b(y)$, only now the outer maxima are
used, as illustrated in \figref{rootbyouter} for the lowest four
convergents. The $b_{[n]}$ of the outer shearless orbits are marked by
solid circles ($\bullet$).\footnote{We use only the smooth maxima on
any symmetry line, since they are numerically easier to find, and the
orbits producing the ``spiky'' maxima have smooth ones on other
symmetry lines. In fact, using only the smooth maxima on $s_1$ to
locate the ``up'' orbits at $y_u$ immediately gives, by applying the
symmetry $S$ the ``down'' orbits on $s_2$ as smooth maxima at $y_d$=$-y_u$.}
As in the central meander case, the best approximation to
$b_\infty(a)$ (here $b_{[34]}$) is used for applying Greene's residue
criterion. Finally, varying $a$ to find the transition between residue
convergence to 0 and $\infty$ results in the parameter values
($a_{co}$, $b_{co}$).

\begin{figure}[t!] \begin{center}
\includegraphics[width=0.45\textwidth]{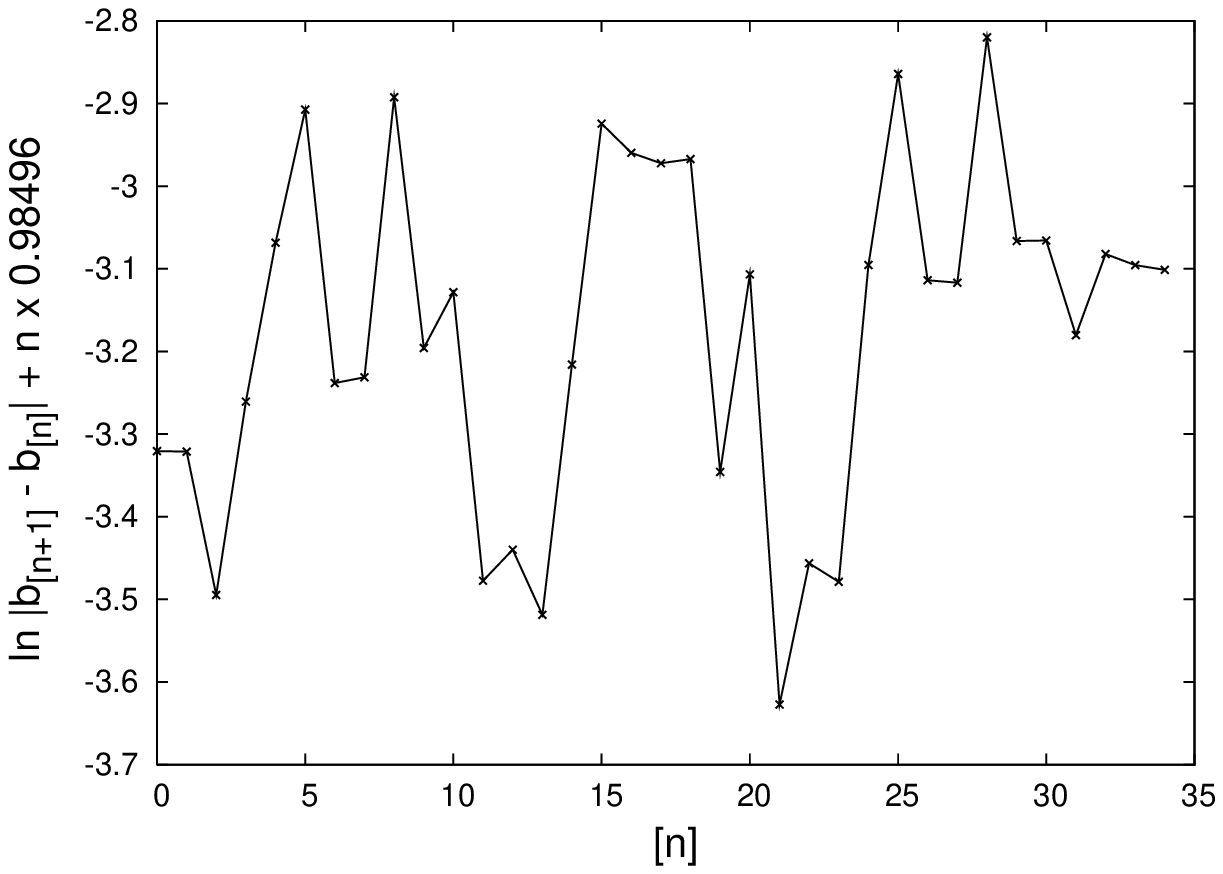}
\hfill
\includegraphics[width=0.45\textwidth]{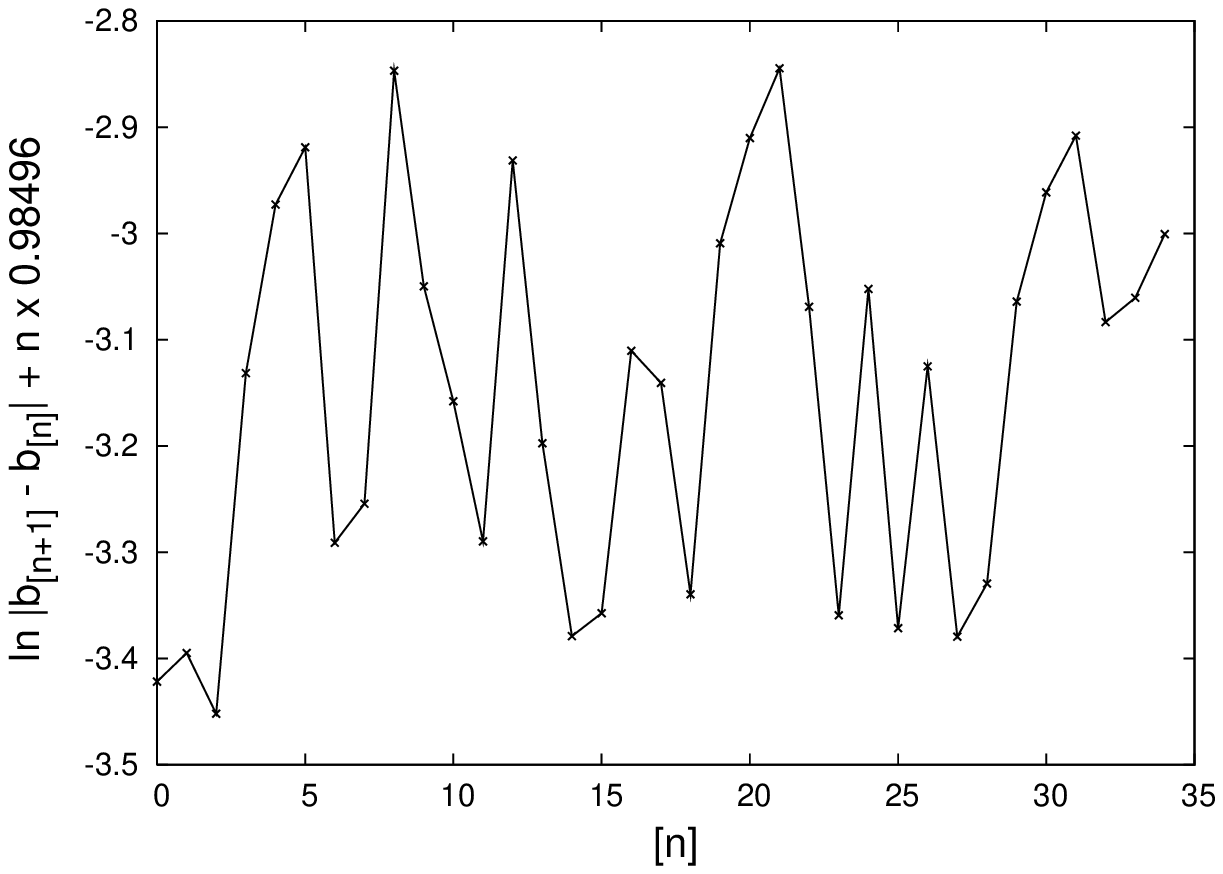}
\caption{$b_{[n]}$ differences in approximating the critical shearless
  $\omega=[0,1,11,1,1,\ldots]$ outer orbits at $a_{co}=0.9757564461$
  on the $s_1$ (above) and $s_4$ (below) symmetry lines.}
\label{fig:bdiffouter} \end{center} \end{figure}

Even though the $b_{[n]}(a)$ seem to follow the same scaling law,
\eqeqref{bscalinglaw}, as the ones approximating the central shearless
meander, the data does not show sufficient evidence for a periodicity of
$B(n)$. Using the numerical value for $\delta_1$ from \eqeqref{delta1a} to
allow for a direct comparison, and again plotting the offsets of the
logarithmic $b_{[n]}$ differences about this average slope,
\figref{bdiffouter} is obtained.\footnote{Since in contrast to the central
meander case, the results on $s_1$ and $s_4$ are not equivalent up to a shift
in $[n] \rightarrow [n+6]$, results for the up orbits on both symmetry lines
are shown separately.  For the down orbits on $s_2$ and $s_3$, the same plots
are found due to the $S$ symmetry.}  Without an apparent periodicity, a better
approximation to $b_\infty$ than $b_{[34]}$ , similar to the extrapolation in
\eqeqref{extrapolation}, cannot be found.

%

\subsection{Residues at breakup}
\label{ssec:nosixc}

\begin{figure*}[t!] \begin{center}
\includegraphics[width=3in]{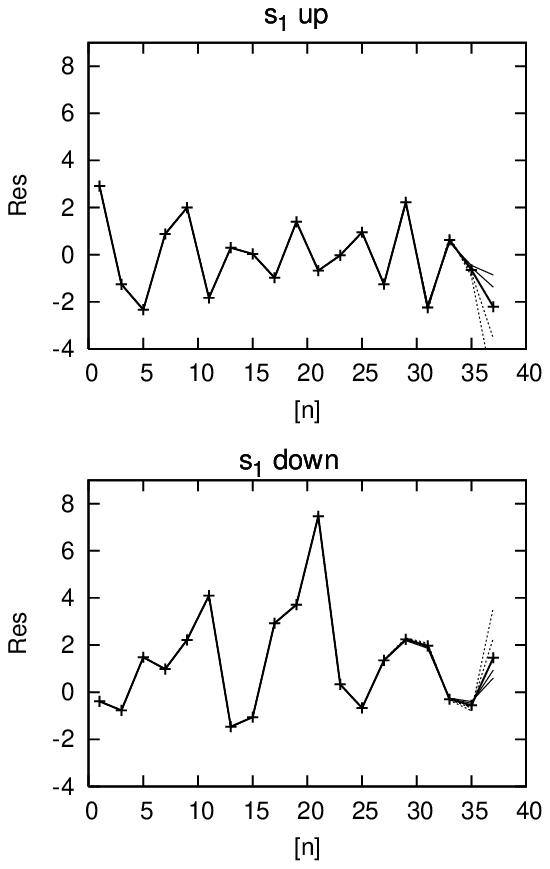}
\includegraphics[width=3in]{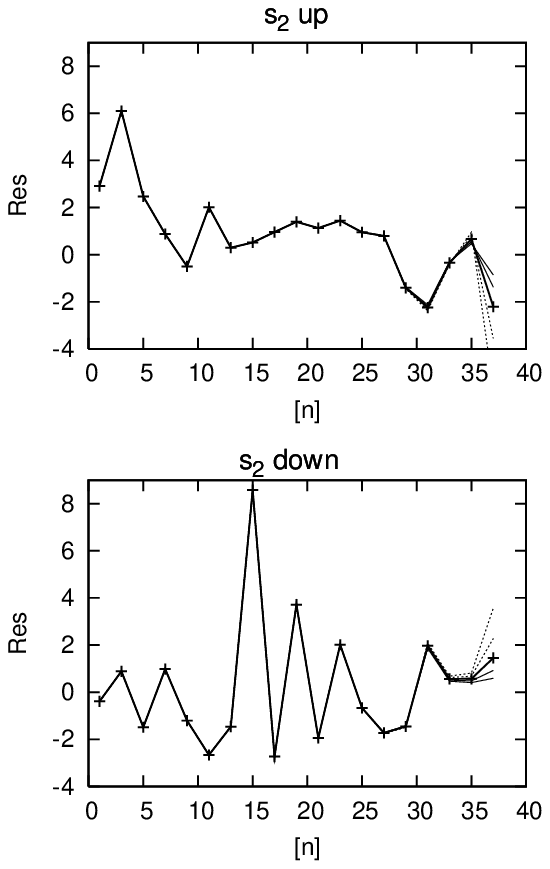}
\includegraphics[width=3in]{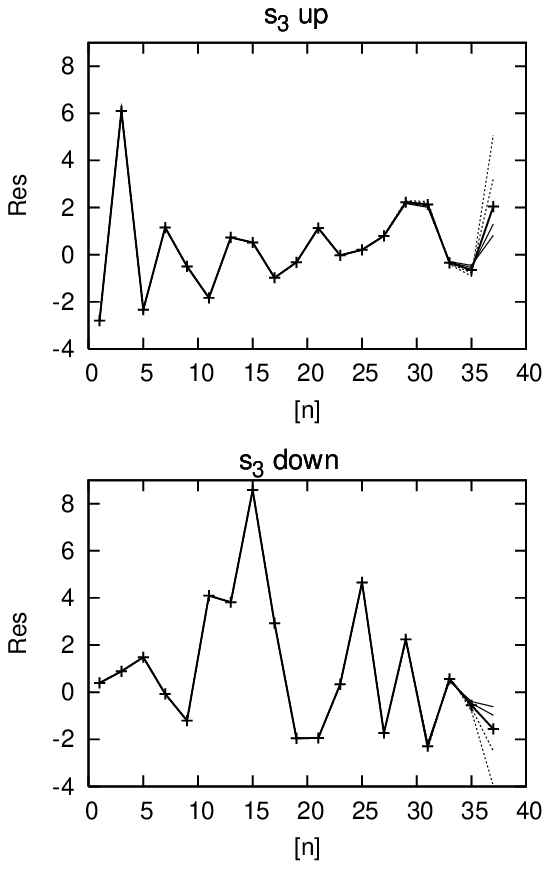}
\includegraphics[width=3in]{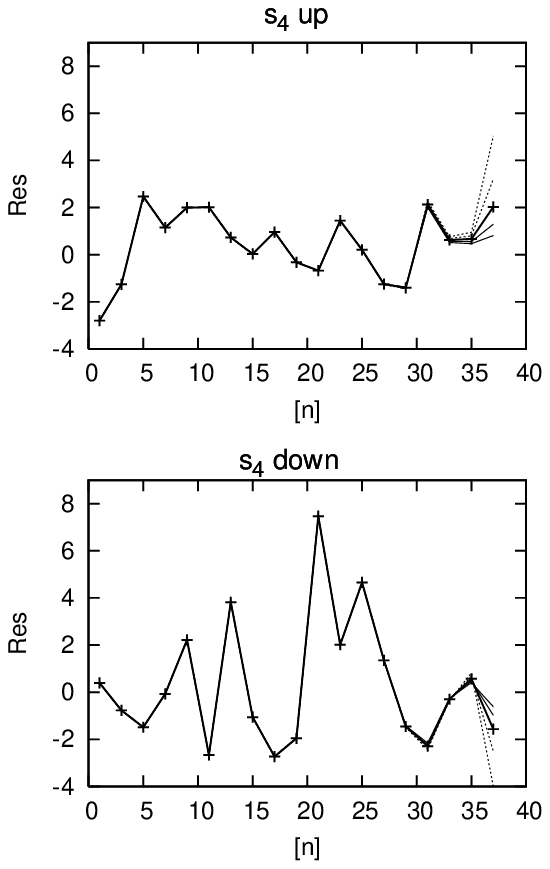}
 \caption{Residue behavior for orbits on $s_1$ to $s_4$ at the breakup
  of $\omega=[0,1,11,1,1,\ldots]$ outer shearless tori. Residues for
  assumed breakup parameter values $(a_{co}, b_{co})$ are indicated by
  the bold line. Residues for five parameter values with $a<a_{co}$ in
  steps of $10^{-10}$ are indicated by thin lines and residues for
  five parameter values with $a>a_{co}$ in steps of $10^{-10}$ by
  dotted lines.}
\label{fig:reso1234} \end{center} \end{figure*}

As before, we search along $\left( a,b_{[34]}(a) \right)$ for the transition
between residue convergence to 0 and $\infty$. However, unlike before, where
the critical parameters could be determined very accurately by observing
significant deviations from the residue six-cycle for small changes in
parameters, here the determination of the exact point of breakup is slightly
less accurate.  Because no clear cycle pattern is apparent, the only criterion
to rely on is the residue convergence to 0 and $\infty$, which, however,
leaves a transition range where the convergence of numerical data is
inconclusive. With this in mind, our ``best guess'' for the critical
parameters for the outer shearless tori breakup is:
\begin{equation}
\label{eq:acbco}
(a_{co}, b_{co})=(0.9757564461, 0.1878717476259388)\;,
\end{equation}
where the last digit in $a_{co}$ and several of the last digits in
$b_{co}$ are merely conjectured to be the values that produce the
``best transitional behavior'' in the residue plots in
\figref{reso1234}.  For a better view of the accuracy of these
critical parameters, all parameters used are listed in
\tabref{critabo}.

\begin{table}[t!]
\begin{center}
\begin{tabular}{|c|c||} \hline
$a_{[34]}$ & $b_{[34]}$ \\
\hline \hline
0.9757564459 & 0.1878717467093720 \\
0.9757564460 & 0.1878717471676554 \\
{\it 0.9757564461} & {\it 0.1878717476259388} \\
0.9757564462 & 0.1878717480842221 \\
0.9757564463 & 0.1878717485425054 \\
\hline
\end{tabular}
\end{center}
\caption{Numerical values of the critical parameters
  $(a_{[34]},b_{[34]})$ in the vicinity of the shearless outer tori
  breakup, with the conjectured critical parameters shown in {\it
  italics}.}
\label{tab:critabo}
\end{table}

In this case, the only (pairs of) existing periodic orbits $[n]$ of
\tabref{cfs} are the ones with odd $n$. Since the outer shearless
orbits (each considered separately) are not $S$-invariant, a residue
correspondence scheme as in \tabref{6cscheme} does not exist.
Therefore residues for both orbits on all four symmetry lines are
shown in \figref{reso1234}.

In contrast to all previous studies of breakups of central shearless
tori with noble winding number, for the outer, non-$S$-invariant tori,
no six-cycles occur. Whether the residues converge to higher period
cycles remains an open question, since within the limited number of 19
data points that could be obtained numerically with sufficient
accuracy (up to $[n]=[37]$ with only odd $[n]$ existing) no such cycle
could be clearly identified, but certainly not ruled out either. What
can be established, however, is that the outer shearless tori
represent the first example where a breakup type different from a
residue six-cycle is observed in the standard nontwist map.

\subsection{Phase space at breakup}
\label{ssec:noscaling}

\begin{figure}[t!] \begin{center}
\includegraphics[width=3in]{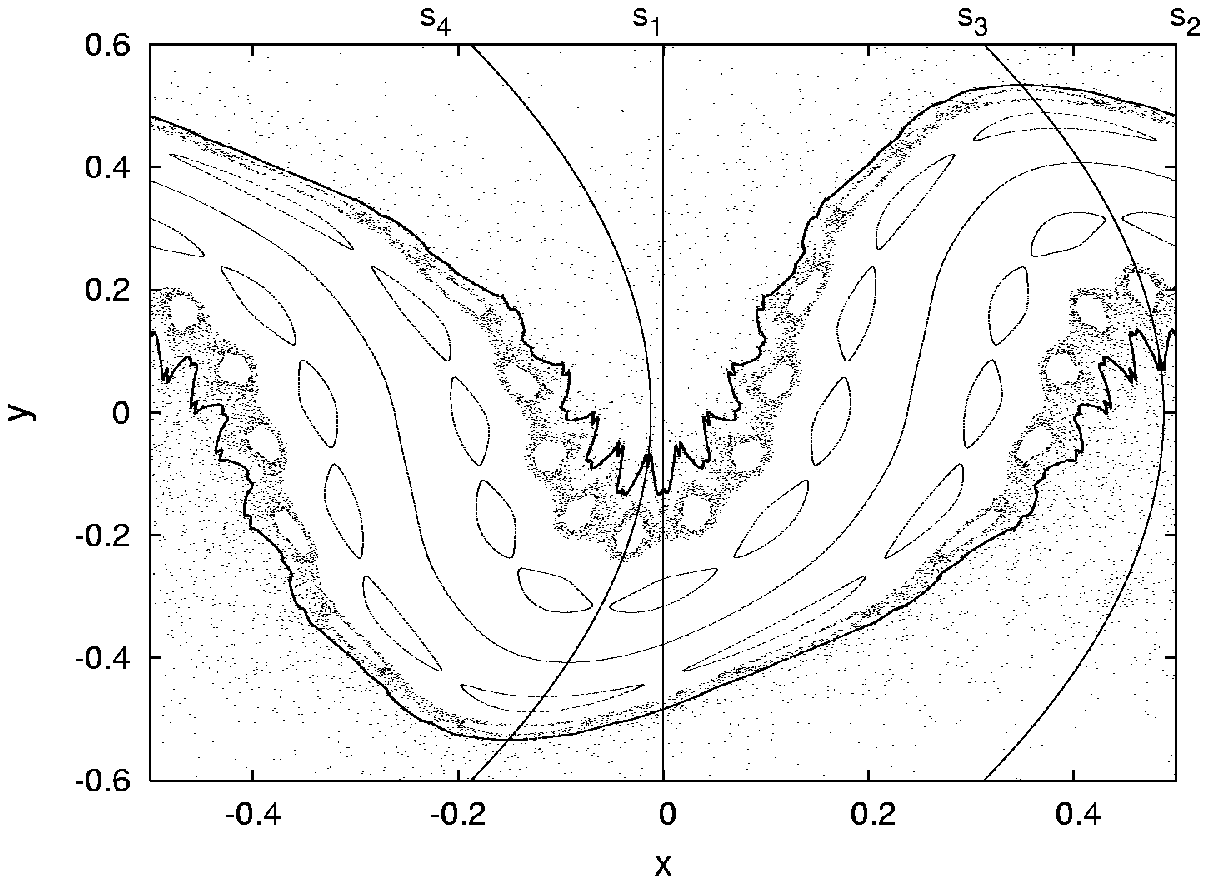}
\caption{ Phase space of the $[0,1,11,1,1,1\ldots]$ outer shearless
  tori at breakup, where $a_{co}= 0.9757564461$ and
  $b_{co}=0.1878717476259388$.  The symmetry lines $s_1$ to $s_4$ are
  shown.}
\label{fig:outer-phsp} \end{center} \end{figure}

The phase space at the critical parameter values for the breakup of
the outer shearless tori is shown in \figref{outer-phsp}. In contrast
to the case of the central meander no scaling invariance was found.

%

\section{Regularity of critical invariant tori}
\label{sec:regularity}

\begin{figure}[t!] \begin{center}
\includegraphics[width=0.45\textwidth]{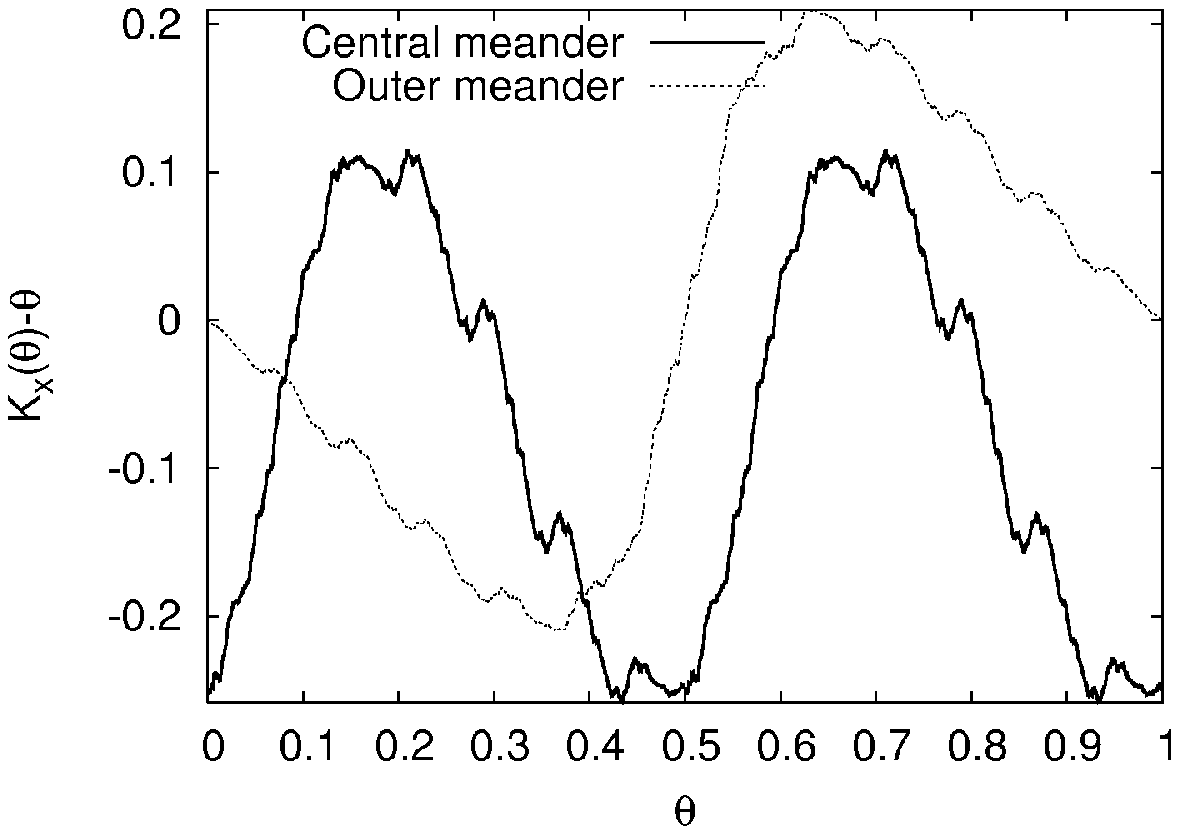}
\includegraphics[width=0.45\textwidth]{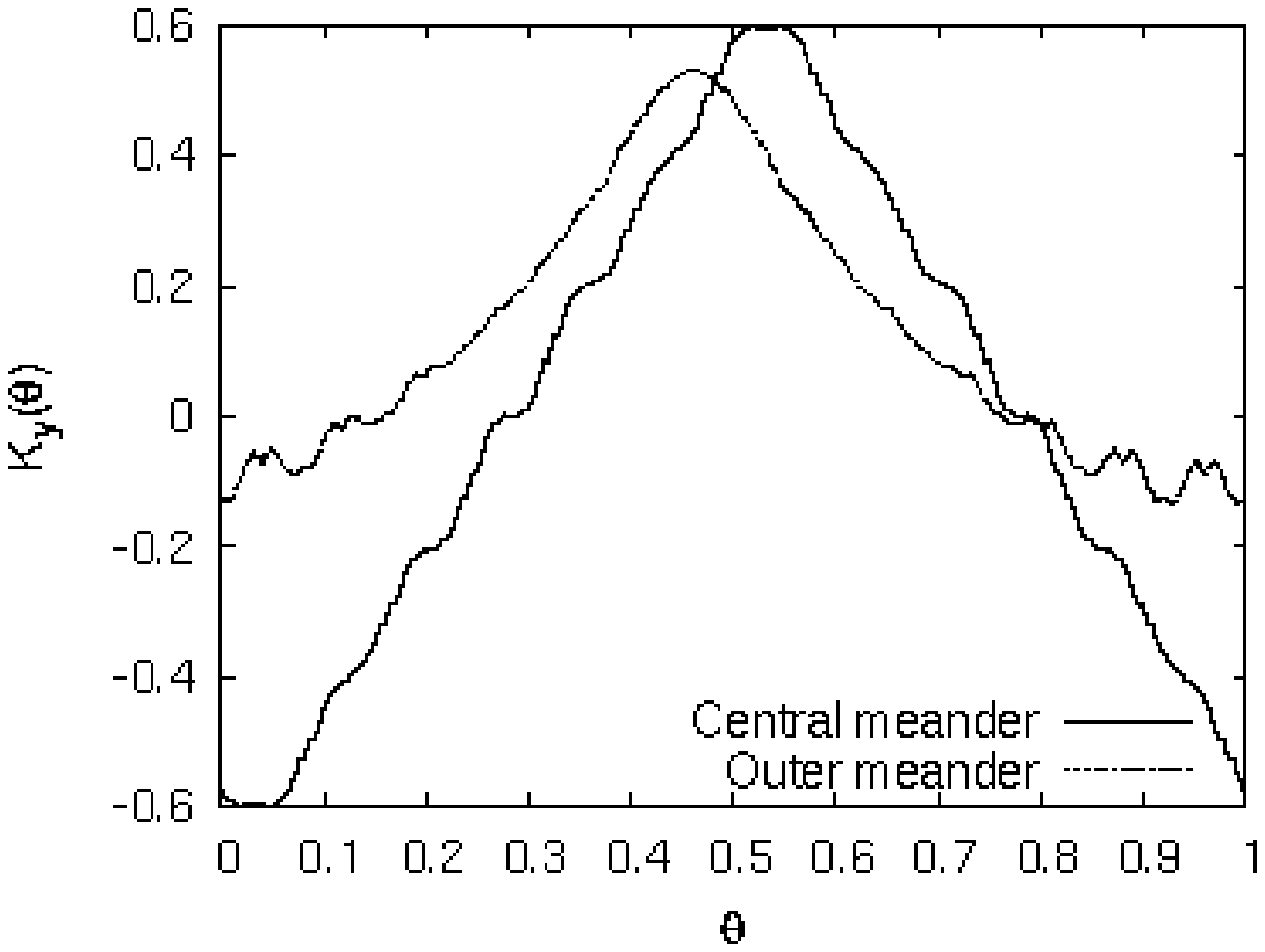}
\caption{The hull functions $K_x(\theta)-\theta$ (left) and
  $K_y(\theta)$ (right) for the critical central and outer meanders.}
\label{fig:hull} \end{center} \end{figure}

The shearless irrational orbits can be described by a function called
the ``hull function.'' For an invariant torus of rotation number
$\omega$, this is given by the map $K : \Tset \to \Tset \times \Rset$
such that $M \circ K(\theta) = K(\theta + \omega),$ and the range of
$K$ is the invariant torus under consideration. We choose the lift
$\tilde{K} : \Rset \to \Rset \times \Rset$ of the map $K$ to satisfy
\begin{eqnarray}
\tilde{K}(\theta+1) = \tilde{K}(\theta) + (1,0)\,,
\label{eq:Kper} \end{eqnarray}
which corresponds to the lift $\tilde{M}$ of the SNM satisfying
$\tilde{M}(x+1,y) = \tilde{M}(x,y) + (1,0).$ Equation \eqref{eq:Kper}
also implies that the functions $\tilde{K}_x(\theta)-\theta$ and
$\tilde{K}_y(\theta)$ are periodic.  These functions for the critical
central and outer meanders are shown in Fig.~\ref{fig:hull}.

\begin{figure*}[t!] \begin{center}
\includegraphics[width=0.45\textwidth]{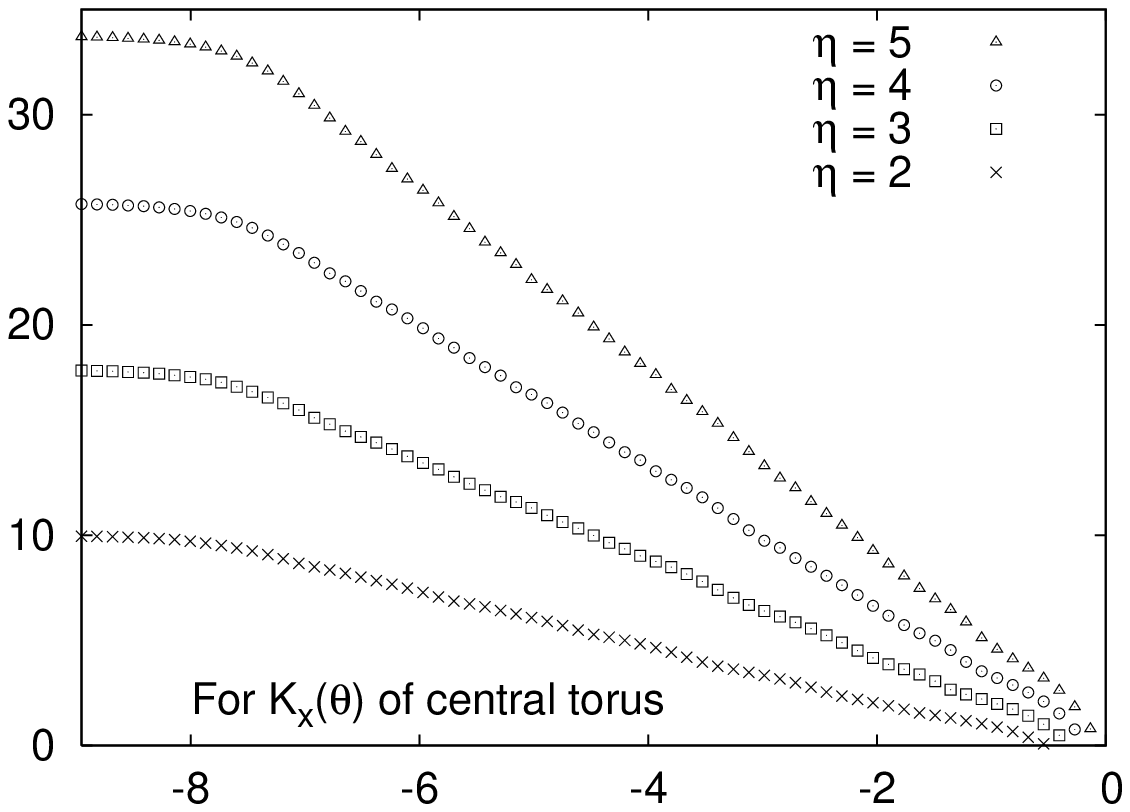}
\includegraphics[width=0.45\textwidth]{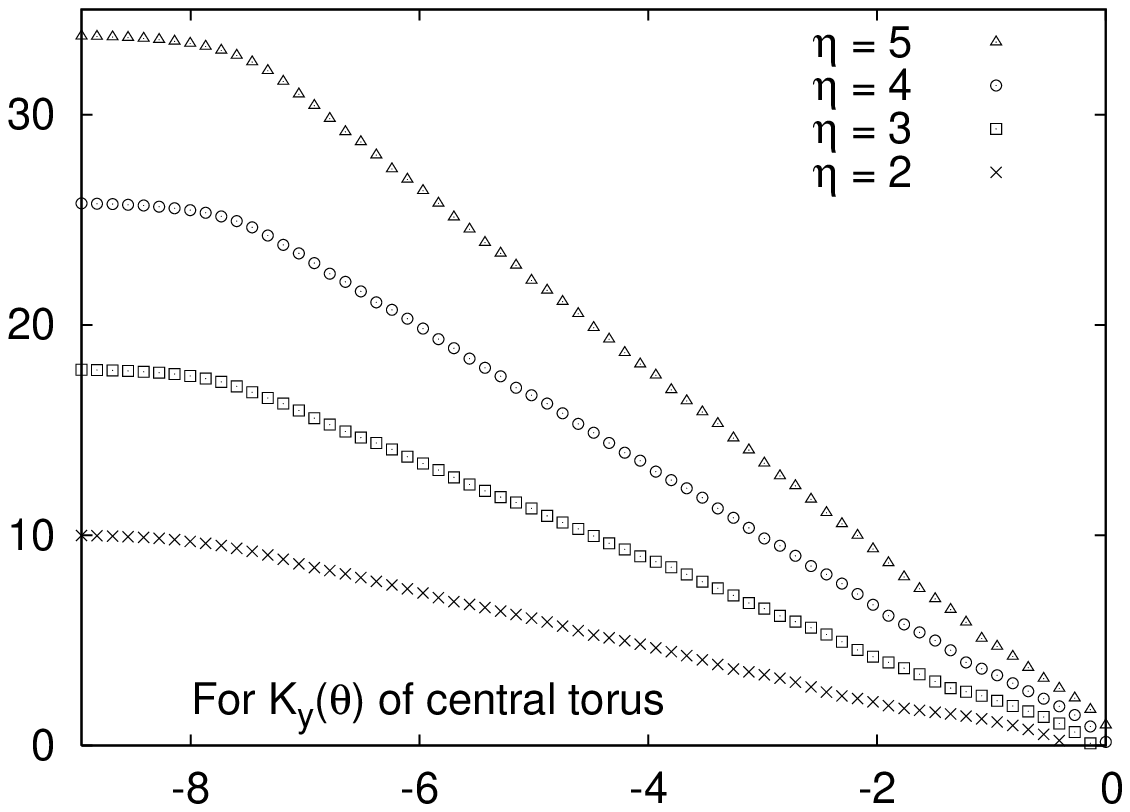}
\includegraphics[width=0.45\textwidth]{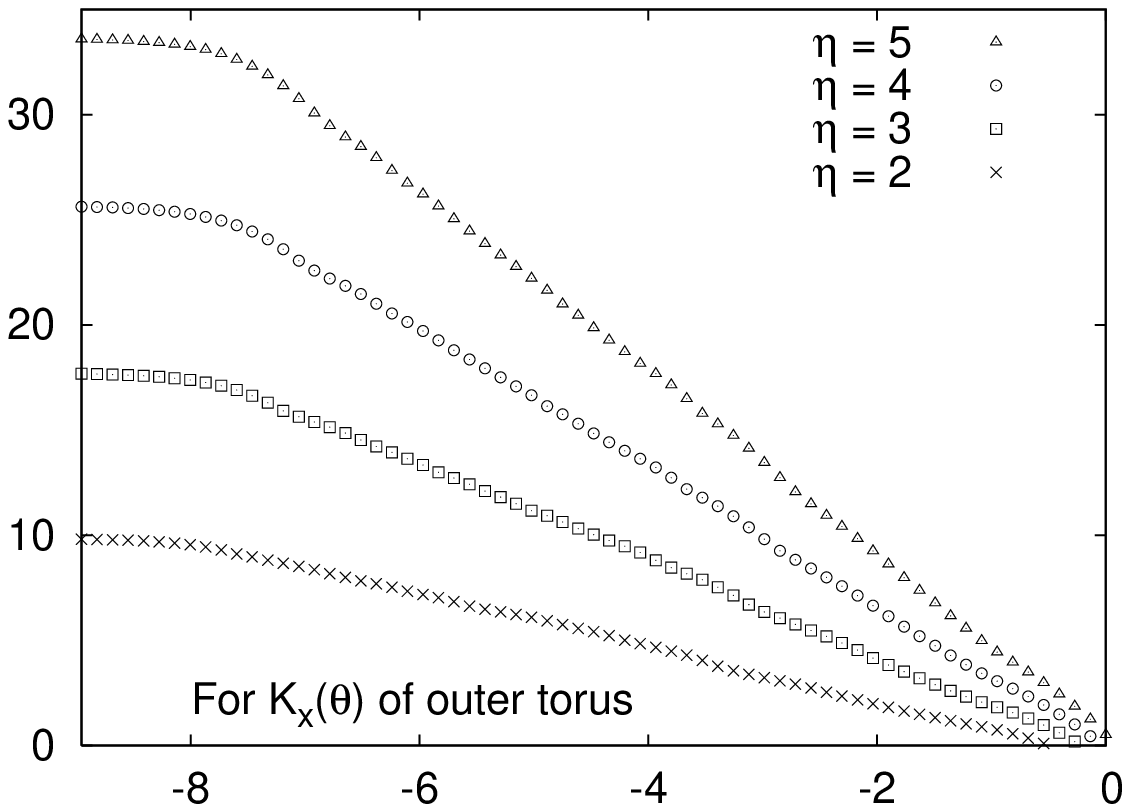}
\includegraphics[width=0.45\textwidth]{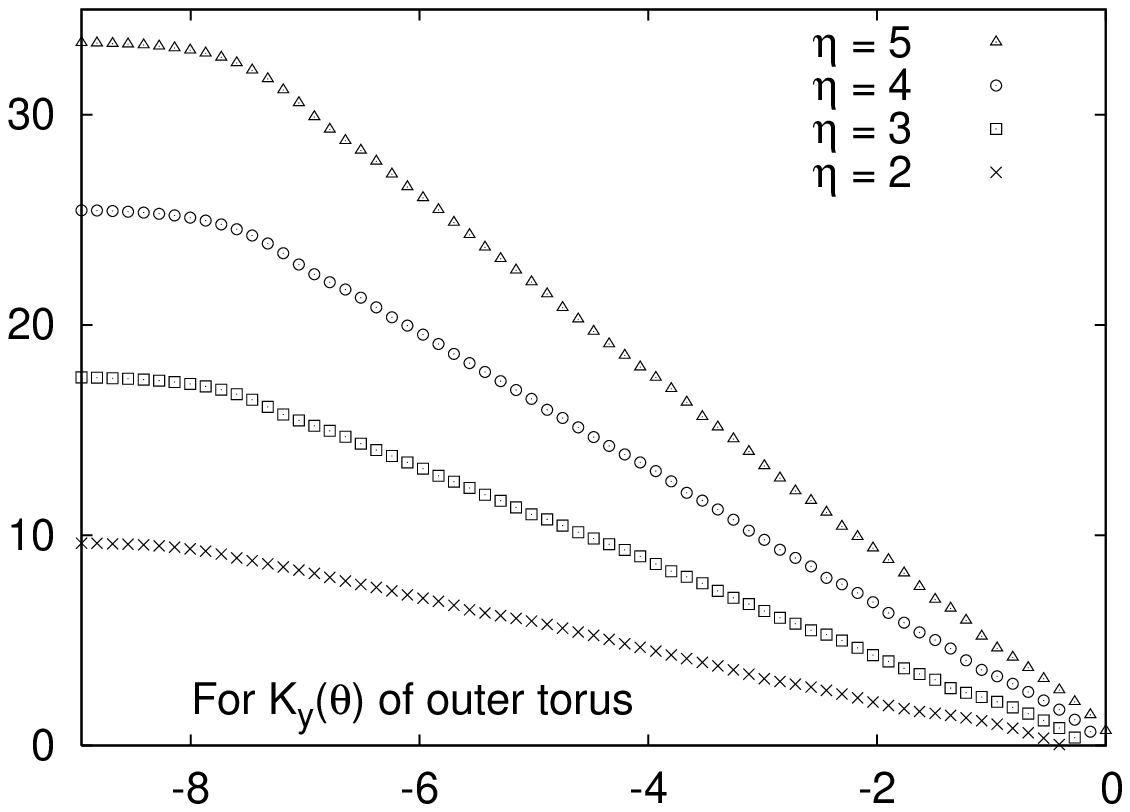}
\caption{The plot of $\log \|(\partial/\partial t)^\eta
  e^{-t\sqrt{-d^2/d\theta^2}} K(\theta)\|_{L^\infty}$ versus $\log t$
  for the hull functions.}
\label{fig:clp} \end{center} \end{figure*}

We studied these functions using techniques from harmonic analysis
developed in \rcites{llave02,apte05b}.  In particular,
Fig.~\ref{fig:clp} shows the plot of $\log \|(\partial/\partial
t)^\eta e^{-t\sqrt{-d^2/d\theta^2}} K(\theta)\|_{L^\infty}$ versus
$\log t$, calculated using $2^{25}$ Fourier coefficients of
$K(\theta)$.

We see that these functions saturate the bounds given in
Ref.~\onlinecite{Stein70}, Ch.~5, Lemma~5: $\|(\partial/\partial
t)^\eta e^{-t\sqrt{-d^2/d\theta^2}} K(\theta)\|_{L^\infty} \le
Ct^{\alpha-\eta}$, where $\alpha$ is the Holder exponent of $K$.  This
allows us to calculate the regularity of these hull functions from the
slope of lines in Fig.~\ref{fig:clp} and the results are presented in
Table~\ref{tab:reg}. We conclude that the regularity of the central
shearless torus is $0.68 \pm 0.02 $ while that of the outer torus is
$0.72 \pm 0.05$.

\begin{table}[t!] \begin{center}
\begin{tabular}{|c||c|c|}\hline
$\eta$ & Central torus     & Outer torus       \\
\hline
 2     & $0.670 \pm 0.002$ & $0.691 \pm 0.004$ \\
 3     & $0.669 \pm 0.004$ & $0.683 \pm 0.005$ \\
 4     & $0.670 \pm 0.004$ & $0.685 \pm 0.006$ \\
 5     & $0.671 \pm 0.004$ & $0.665 \pm 0.006$ \\
\hline
$K_x(\theta)$ & $0.67 \pm 0.01$ & $0.69 \pm 0.01$ \\
\hline
\hline
$\eta$ & Central torus     & Outer torus       \\
\hline
 2     & $0.692 \pm 0.003$ & $0.756 \pm 0.003$ \\
 3     & $0.696 \pm 0.003$ & $0.757 \pm 0.003$ \\
 4     & $0.693 \pm 0.003$ & $0.769 \pm 0.004$ \\
 5     & $0.694 \pm 0.004$ & $0.757 \pm 0.004$ \\
\hline
$K_y(\theta)$ & $0.69 \pm 0.01$ & $0.76 \pm 0.01$ \\
\hline
\end{tabular} \label{tab:reg}
\caption{The regularities of hull functions $K_x(\theta)-\theta$ (top)
  and $K_y(\theta)$(bottom), found using the slopes of lines in
  Fig.~\ref{fig:clp}}
\end{center} \end{table}
This agrees very well with the regularity of other shearless noble
tori studied in \rcite{apte05b}.

\section{Conclusion}
\label{sec:conclusion}

In this paper we presented the breakup of two types of shearless
invariant tori with noble winding number that had not been studied
previously: a central meander and an outer torus. The breakup of the
central meander showed within numerical accuracy the same critical
residues, scaling parameters, and eigenvalues of the renormalization
group operator as the central shearless invariant tori previously
studied.  From a renormalization group point of view this was to be
expected: all nontwist maps with a critical shearless torus of noble
winding number are expected to be equivalent under renormalization to
the map with the critical shearless golden mean torus, independent of
being a meander or not.

In this light, the result of the outer torus breakup is surprising.
Although the winding number is noble, no critical residue pattern
could be established within the numerically accessible range. This
suggests that the number theoretic properties of the winding number
might not be enough for the classification of different breakup
scenarios. In the case of nontwist maps the symmetry properties of the
shearless torus (here: $S$-invariant vs. not $S$-invariant) seem to
affect the breakup.
It is possible that after an appropriate coordinate change, that will
make the outer torus symmetric in those coordinates, the SNM with
critical outer torus is equivalent under renormalization to the fixed
point with critical shearless golden mean torus.  Alternatively, this
could be the first indication of a new fixed point of the
renormalization group operator for area-preserving maps.

\section*{Acknowledgments}

This research was supported by US DOE Contract DE-FG01-96ER-54346.  AW
thanks the Dept.\ of Physical and Biological Sciences at WNEC for
travel support.

\appendix

\section{Basic definitions}
\label{sec:basics}

For reference, we list a few basic definitions used throughout the
main text:

An {\it orbit} of an area-preserving map $M$ is a sequence of points
$\left\{\left(x_i,y_i\right)\right\}_{i=-\infty}^{\infty}$ such that
$M\left(x_i,y_i\right) = \left(x_{i+1},y_{i+1}\right)$. The {\it
winding number} $\omega$ of an orbit is defined as the limit $\omega =
\lim_{i\to\infty} (x_i/i)$, when it exists. Here the $x$-coordinate is
``lifted'' from $\Tset$ to $\Rset$. A {\it periodic orbit} of period
$n$ is an orbit $M^n \left( x_i, y_i\right) = \left( x_i+m,
y_i\right)$, $\forall \:i$, where $m$ is an integer. Periodic orbits
have rational winding numbers $\omega=m/n$. An {\it invariant torus}
is a one-dimensional set $C$, a curve, that is invariant under the
map, $C = M(C)$. Of particular importance are the invariant tori that
are homeomorphic to a circle and wind around the $x$-domain because,
in two-dimensional maps, they act as transport barriers. Orbits
belonging to such a torus generically have irrational winding number.

A map $M$ is called {\it reversible} if it can be decomposed as
$M=I_1\circ I_2$ with $I_i^2=0$. The fixed point sets of $I_i$ are
one-dimensional sets, called the {\it symmetry lines} of the map.  For
the SNM the symmetry lines are $s_1=\{(x,y)|x=0\}$,
$s_2=\{(x,y)|x=1/2\}$, $s_3=\{(x,y)|x=a\left(1-y^2\right)/2\}$, and
$s_4=\{(x,y)|x=a\left(1-y^2\right)/2+1/2\}$.

The $m/n$-bifurcation curve $b=\Phi_{m/n,i}(a)$ is the set of $(a,b)$
values for which the $m/n$ up and down periodic orbits on the symmetry
line $s_i$ are at the point of collision. The main property of this
curve is that for $(a,b)$ values below $b=\Phi_{m/n,i}(a)$, the $r/s$
periodic orbits with $r/s<m/n$ exist.  Thus, $m/n$ is the maximum
winding number for parameter values along the $m/n$ bifurcation
curve. As detailed in~\rcite{wurm05}, in certain parameter regions
multiple orbits of winding number $m/n$, and therefore multiple
bifurcation curves of the same winding number, can exist.

\end{document}